\documentclass[longauth]{aa} 
\usepackage[authoryear]{natbib}
\usepackage{graphicx}

\begin{document}
   \title{Optical configuration and analysis of the AMBER/VLTI instrument}
%

%
\newcommand{\luan}{1}
\newcommand{\gemini}{2}
\newcommand{\oaa}{3}
\newcommand{\mpifr}{4}
\newcommand{\laog}{5}
\newcommand{\esoparanal}{6}
\newcommand{\dtinsu}{7}
\newcommand{\onera}{8}
\newcommand{\cral}{9}
\newcommand{\ircom}{10}
\newcommand{\esogarching}{11}
\newcommand{\kis}{12}
\newcommand{\acke}{13}
\newcommand{\gil}{14}
\newcommand{\mouillet}{15}

\author{%
       S.~Robbe-Dubois\inst{\luan}              
  \and S.~Lagarde\inst{\gemini}                    
  \and R.G.~Petrov\inst{\luan}                  
  \and F.~Lisi\inst{\oaa}                       
  \and U.~Beckmann\inst{\mpifr}                 
  \and P.~Antonelli\inst{\gemini}                  
  \and Y.~Bresson\inst{\gemini}                    
  \and G.~Martinot-Lagarde\inst{\gemini,\dtinsu}                    
  \and A.~Roussel\inst{\gemini}
  \and P.~Salinari\inst{\oaa}                             
  \and M.~Vannier\inst{\luan,\esoparanal}                            	  
  \and A.~Chelli\inst{\laog}                  
  \and M.~Dugu\'e\inst{\gemini}                  
  \and G.~Duvert\inst{\laog}                    
  \and S.~Gennari\inst{\oaa}                           	  
  \and L.~Gl\"uck\inst{\laog}                   
  \and P.~Kern\inst{\laog}                      
  \and E.~Le Coarer\inst{\laog}                 
  \and F.~Malbet\inst{\laog}
  \and F.~Millour\inst{\luan,\laog}             
 	\and K.~Perraut\inst{\laog}       
 	\and P.~Puget\inst{\laog}                     
  \and F.~Rantakyr\"o\inst{\esoparanal}              
  \and E.~Tatulli\inst{\laog}                                 
  \and G.~Weigelt\inst{\mpifr}                    
  \and G.~Zins\inst{\laog}                        
  \and M.~Accardo\inst{\oaa}                                    
  \and B.~Acke\inst{\laog,\acke}                                   
  \and K.~Agabi\inst{\luan}                                   	  
  \and E.~Altariba\inst{\laog}                         
  \and B.~Arezki\inst{\laog}                                            
  \and E.~Aristidi\inst{\luan}                                  
  \and C.~Baffa\inst{\oaa}                                        
  \and J.~Behrend\inst{\mpifr}                                          
  \and T.~Bl\"ocker\inst{\mpifr}                                  
  \and S.~Bonhomme\inst{\gemini}                                           
  \and S.~Busoni\inst{\oaa}                                       
  \and F.~Cassaing\inst{\onera}                                              
  \and J.-M.~Clausse\inst{\gemini}                        	          
  \and J.~Colin\inst{\gemini}		          
  \and C.~Connot\inst{\mpifr}                          
  \and L.~Delage\inst{\ircom}                            	            
  \and A.~Delboulb\'e\inst{\laog}                      
  \and A.~Domiciano de Souza\inst{\luan,\gemini}              
  \and T.~Driebe\inst{\mpifr}                          	    
  \and P.~Feautrier\inst{\laog}                        		  
  \and D.~Ferruzzi\inst{\oaa}                          	  
  \and T.~Forveille\inst{\laog}                        		  
  \and E.~Fossat\inst{\luan}                           	          
  \and R.~Foy\inst{\cral}                                  		  
  \and D.~Fraix-Burnet\inst{\laog} 
  \and A.~Gallardo\inst{\laog}                         	  
  \and E.~Giani\inst{\oaa}                             		  
  \and C.~Gil\inst{\laog,\gil}                           		  
  \and A.~Glentzlin\inst{\gemini}                         		  
  \and M.~Heiden\inst{\mpifr}                          	          
  \and M.~Heininger\inst{\mpifr}                       		  
  \and O.~Hernandez Utrera\inst{\laog}                   
  \and K.-H.~Hofmann\inst{\mpifr}                        
  \and D.~Kamm\inst{\gemini}                                
  \and M.~Kiekebusch\inst{\esoparanal}                                     
  \and S.~Kraus\inst{\mpifr}                         
  \and D.~Le Contel\inst{\gemini}                         
  \and J.-M.~Le Contel\inst{\gemini}                      
  \and T.~Lesourd\inst{\dtinsu}
  \and B.~Lopez\inst{\gemini}                               
  \and M.~Lopez\inst{\dtinsu}
  \and Y.~Magnard\inst{\laog}                            
  \and A.~Marconi\inst{\oaa}                           
  \and G.~Mars\inst{\gemini}                              
  \and P.~Mathias\inst{\gemini}                           
  \and P.~M\`ege\inst{\laog}                           
  \and J.-L.~Monin\inst{\laog}                           
  \and D.~Mouillet\inst{\laog,\mouillet}                      
  \and D.~Mourard\inst{\gemini}                             
  \and E.~Nussbaum\inst{\mpifr}                          
  \and K.~Ohnaka\inst{\mpifr}                        
  \and J.~Pacheco\inst{\gemini}		          
  \and C.~Perrier\inst{\laog}                        	            
  \and Y.~Rabbia\inst{\gemini}                                          
  \and S.~Rebattu\inst{\gemini}                         	          
  \and F.~Reynaud\inst{\ircom}                            	            
  \and A.~Richichi\inst{\esogarching}                          	            
  \and A.~Robini\inst{\luan}
  \and M.~Sacchettini\inst{\laog}                           	  
  \and D.~Schertl\inst{\mpifr}                                          
  \and M.~Sch\"oller\inst{\esoparanal}
  \and W.~Solscheid\inst{\mpifr}          
  \and A.~Spang\inst{\gemini}                      
  \and P.~Stee\inst{\gemini}                                       
  \and P.~Stefanini\inst{\oaa}                                          
  \and M.~Tallon\inst{\cral}                             	    
  \and I.~Tallon-Bosc\inst{\cral}                               	    
  \and D.~Tasso\inst{\gemini}                                    
  \and L.~Testi\inst{\oaa}                                              
  \and F.~Vakili\inst{\luan}
  \and O.~von der L\"uhe\inst{\kis}                                         
  \and J.-C.~Valtier\inst{\gemini}
  \and N.~Ventura\inst{\laog}                        		
}             
\institute{
  Laboratoire Universitaire d'Astrophysique de Nice, UMR 6525
  Universit\'e de Nice - Sophia Antipolis/CNRS, Parc Valrose, F-06108
  Nice cedex 2, 
  France
  \and Laboratoire Gemini, UMR 6203 Observatoire de la C\^ote
  d'Azur/CNRS, BP 4229, F-06304 Nice Cedex 4, France
  \and INAF-Osservatorio Astrofisico di Arcetri, Istituto Nazionale di
  Astrofisica, Largo E. Fermi 5, I-50125 Firenze, Italy
  \and Max-Planck-Institut f\"ur Radioastronomie, Auf dem H\"ugel 69,
  D-53121 Bonn, Germany
  \and Laboratoire d'Astrophysique de Grenoble, UMR 5571 Universit\'e Joseph
  Fourier/CNRS, BP 53, F-38041 Grenoble Cedex 9, France
  \and European Southern Observatory, Casilla 19001, Santiago 19,
  Chile
  \and Division Technique INSU/CNRS UPS 855, 1 place Aristide
  Briand, F-92195 Meudon cedex, France
  \and ONERA/DOTA, 29 av de la Division Leclerc, BP 72, F-92322
  Chatillon cedex, France 
  \and Centre de Recherche Astronomique de Lyon, UMR 5574 Universit\'e
  Claude Bernard/CNRS, 9 avenue Charles Andr\'e, F-69561 Saint Genis
  Laval cedex, France
  \and IRCOM, UMR 6615 Universit\'e de Limoges/CNRS, 123 avenue Albert
  Thomas, F-87060 Limoges cedex, France
  \and European Southern Observatory, Karl Schwarzschild Strasse 2,
  D-85748 Garching, Germany
  \and Kiepenheuer Institut f\"ur Sonnenphysik, Sch\"oneckstr. 6,
  D-79104 Freiburg, Germany 
  \and Instituut voor Sterrenkunde, KU-Leuven, Celestijnenlaan 200B,
  B-3001 Leuven, Belgium 
  \and Centro  de  Astrof\'{\i}sica  da  Universidade  do  Porto, Rua
  das Estrelas - 4150-762 Porto, Portugal 
  \and Laboratoire Astrophysique de Toulouse, UMR 5572 Universit\'e
       Paul Sabatier/CNRS, BP 826, F-65008 Tarbes cedex, France 
}

   \offprints{S. Robbe-Dubois, \email{robbe@unice.fr}}  
   \date{Received:/ Accepted:}
   \abstract
   {}
  {This paper describes the design goals and engineering efforts that led to the realization of AMBER (Astronomical Multi BEam combineR) and to the achievement of its present performance.}
  {On the basis of the general instrumental concept, AMBER was decomposed into modules whose functions and detailed characteristics are given. Emphasis is put on the spatial filtering system, a key element of the instrument. We established a budget for transmission and contrast degradation through the different modules, and made the detailed optical design. The latter confirmed the overall performance of the instrument and defined the exact implementation of the AMBER optics.}
  {The performance was assessed with laboratory measurements and commissionings at the VLTI, in terms of spectral coverage and resolution, instrumental contrast higher than 0.80, minimum magnitude of 11 in K, absolute visibility accuracy of 1\%, and differential phase stability of 10$^{-3}$rad over one minute.}
  {}
  
  \maketitle
\section{Introduction}
AMBER \citep{petrovd} is the near infrared focal beam combiner of the Very Large Telescope Interferometric mode (VLTI). A consortium of French, German, and Italian institutes is in charge of the specification, design, construction, integration, and commissioning of AMBER (http://amber.obs.ujf-grenoble.fr). This instrument is designed to combine three beams coming from any combination of Unit 8-m Telescopes (UT) or Auxiliary 1.8-m Telescopes (AT). The spectral coverage is from 1 to 2.4$~\mu$m  with a priority to the K band. 

AMBER is a general user instrument with a very wide range of astrophysical applications, as is shown in \citet{richichi} and \citet{petrovd}. To achieve the ambitious programs, \citet{petrovb} presented the associated specifications and goals, such as: \newline
\indent- Spectral coverage: J, H, and K bands, from 1.0 $\mu$m to 2.3 $\mu$m (goal: up to 2.4 $\mu$m). \newline
\indent- Spectral resolutions: minimum spectral resolution 30$<$R$<$50, medium spectral resolution 500$<$R$<$1\,000, and highest spectral resolution 10\,000$<$R$<$15\,000.\newline
\indent- Minimum magnitude: K=11, H=11 (goals: K=13, H=12.5, J=11.5).\newline
\indent- Absolute visibility accuracy: 3$\sigma_V$=0.01 (goal: $\sigma_V$=10$^{-4}$).\newline
\indent- Differential phase stability: 10$^{-3}$rad (goal: 10$^{-4}$rad) over one minute integration. This allows us to compute phase closure that is necessary in the search of brown dwarfs and extra solar planets \citep{segr}.\newline

These specifications have been the starting point of a global system analysis \citep{malbet} initiated by a group of interferometrists from several French institutes and completed by the Interferometric GRoup (IGR) of AMBER. This work led to the current definition of the AMBER instrument whose broad outlines are recalled in the paragraphs below.

To reach a sufficient sensitivity, ESO provided a 60 actuator curvature sensing system MACAO (Multi-Application Curvature Adaptive Optics) \citep{arseno} specified to deliver at least a 50\% Strehl ratio $@$ 2.2$\mu$m for on-axis bright sources (V=8) under median seeing conditions (0.65") and a 25\% Strehl ratio $@$ 2.2$\mu$m for faint sources (V=15.5) under the same seeing conditions. The required high accuracy of the absolute visibility measurements implies the use of spatial filters \citep{mege,tatub} with single mode fibers based on the experience of other smaller, successful interferometers such as IOTA/FLUOR \citep{coude}. The atmospheric noise is reduced to photometric fluctuations, which can be monitored, and to Optical Path Difference (OPD) fluctuations between the different pupils, which can be frozen by very short exposures or adaptively corrected by a fringe tracker.

The simultaneous observation of different spectral channels is ensured by dispersed fringes. This very significantly increases the number and the quality of the measurements and subsequently the constraints imposed on the atmospherical models. The modularity of the concept was a strong argument in favor of the multi-axial scheme, as carried out on the Grand Interferom\`etre \`a 2 T\'elescopes (GI2T) of the Plateau de Calern \citep{mourard}. In addition, it was demonstrated that the instrument must correct the atmospheric transversal dispersion in J and H \citep{tallon}. The need of an image cold stop was assessed by \citet{malbet1} to reduce the thermal background coming from the blackbody emission of the fiber heads, which can be greater than the
detector RON, especially for long time exposures in the K-Band. A pupil mask also acts as a cold stop. To perform the data reduction, the ABCD algorithm \citep{chelli,millour}, as used with co-axial configurations with a temporal coding, was chosen. The associated complete data calibration procedure was then fully defined \citep{hofman,chelli,tatuc}.

On the basis of the general instrumental concept resulting from the global system analysis, we defined the main modules and necessary accessories (such as the alignment units) of AMBER. We established a budget for throughput and contrast degradation through the different modules, made the detailed optical design, and performed a complete optical analysis. The latter confirmed the expected overall performance of the instrument in terms of signal-to-noise ratio (SNR). This entire process is described in the present paper. The procedure described here to allocate the specifications of the different modules of an interferometer could be used, after some changes, for the design of other interferometrical instruments, such as, for example, the VLTI second generation instruments.

\section{Overview of the AMBER implementation}
\begin{figure}[h]
\includegraphics[width=84mm]{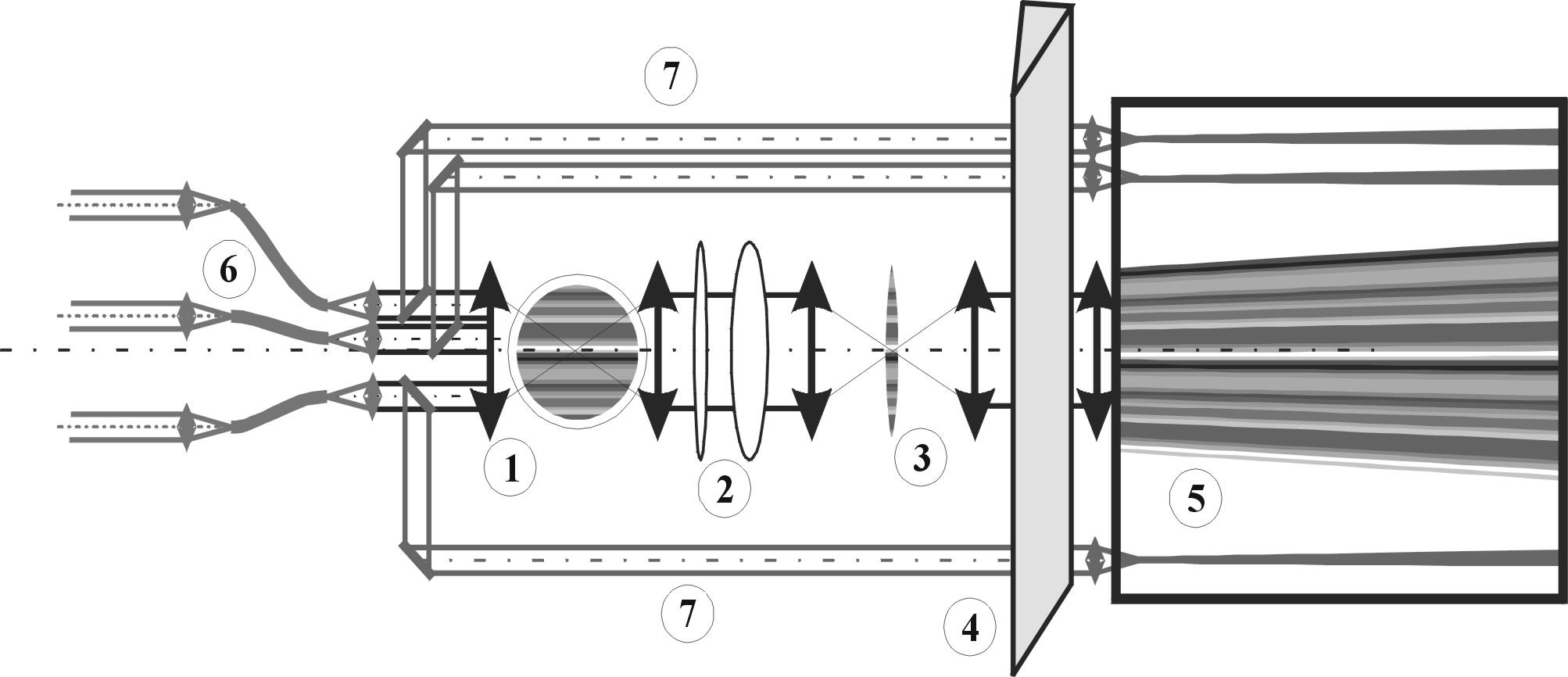}
\caption{Scheme of the AMBER configuration: (1) multiaxial beam combiner; (2) cylindrical optics; (3) anamorphosed focal image with fringes; (4) "long slit spectrograph"; (5) dispersed fringes on 2D detector; (6) spatial filter with single mode optical fibers; (7) photometric beams.}
\label{schem}
\end{figure}

The concept of AMBER is illustrated by Fig.~\ref{schem}. Each input beam is fed into a single mode fiber that reduces all chromatic wavefront perturbations to photometric and global OPD fluctuations (6). At the output of the fibers, the beams are collimated, maintained parallel, and then focused in a common Airy disk (1). The latter contains Young fringes with spacings specific to each baseline, allowing us to separate the interferograms in the Fourier space. This Airy disk goes through the spectrograph slit (3) after being anamorphosed by cylindrical optics (2). The spectrograph (4) forms dispersed fringes on the detector (5), where each column allows us to analyse the interferograms in a different spectral channel. A fraction of each beam is collected before the beam combination to monitor the photometry variations (7).

\begin{figure*}
\begin{center}
\includegraphics[height=60mm]{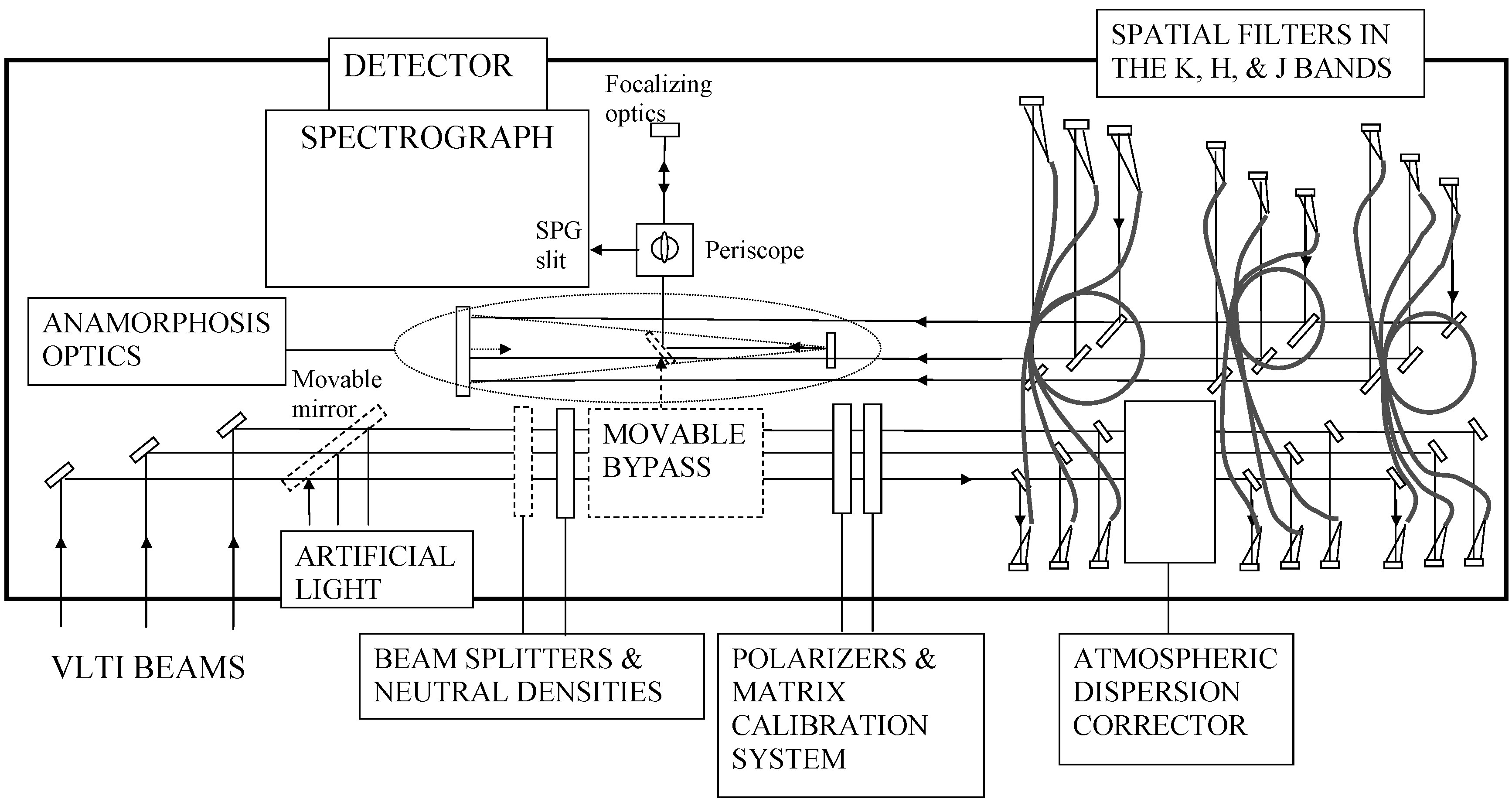}
\includegraphics[height=110mm]{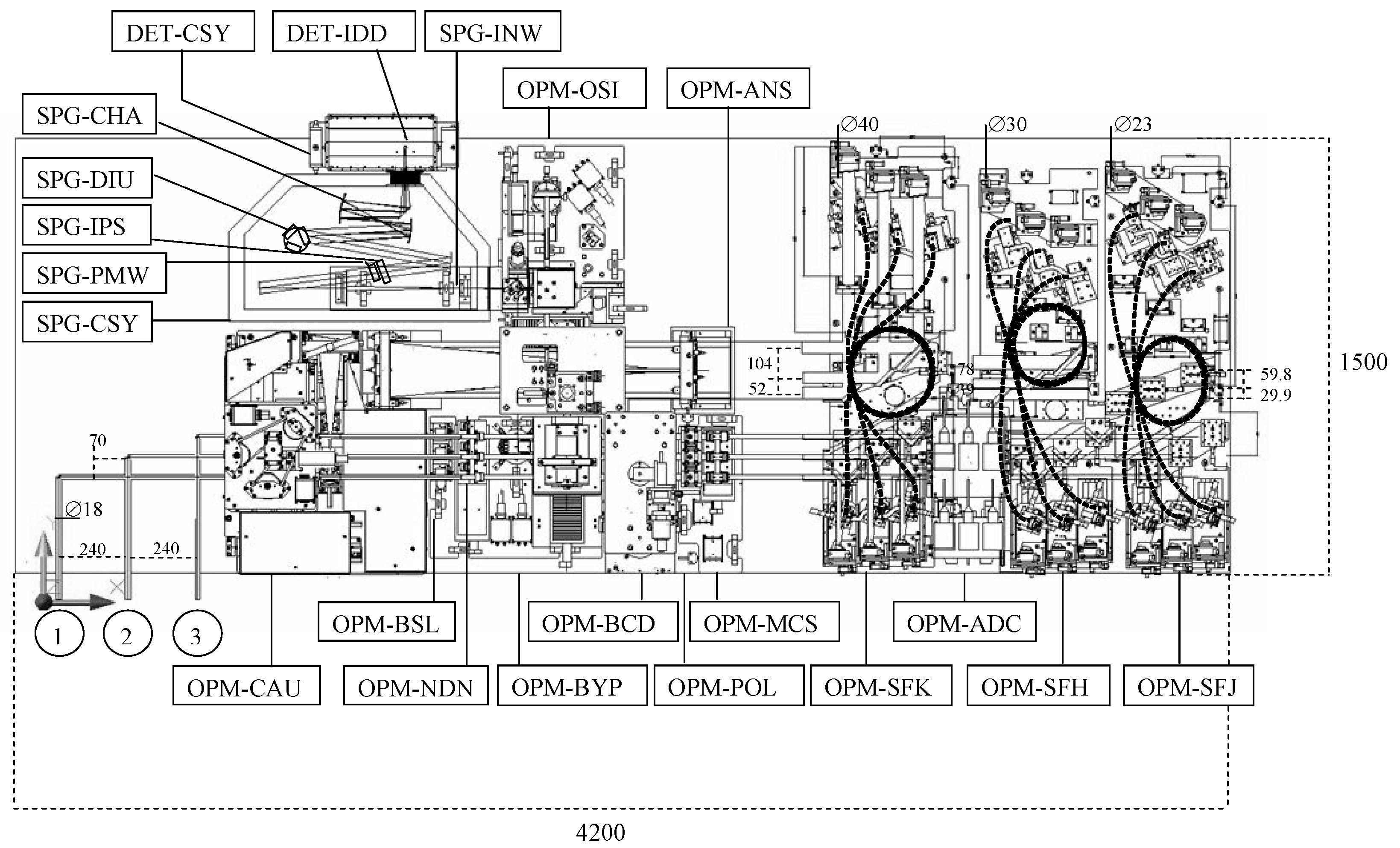}
\end{center}\caption{General implementation of AMBER. The top scheme shows the light path from the VLTI to the detector. Detailed configuration below. The VLTI beams arrive in the lower left corner. OPM-CAU: Calibration and Alignment Unit. OPM-BCD: beam inverting device \citep{petrovc}. OPM-POL: polarization selecting device. OPM-SFK (SFH, SFJ): spatial filters for the K-, H-, and J-bands. OPM-ADC: corrector for the atmospheric differential refraction in H and J. OPM-ANS: cylindrical afocal system for image anamorphosis. OPM-OSI: periscope to co-align the warm and the cold optics. OPM-BYP: movable bypass directly sending the VLTI beams towards technical tools or towards the spectrograph to check VLTI alignment and acquire complex fields. SPG-INW: input wheel with image cold stop and diaphgram inside the spectrograph (SPG). SPG-PMW: pupil mask wheel. SPG-IPS: beam splitter allowing the separation between interferometric and photometric beams. SPG-DIU: light dispersion (gratings or prism). SPG-CHA: SPG camera. DET-IDD: chip. SPG-CSY and DET-CSY: cryostats of the SPG and of the Hawaii detector (DET). During final operation, the two cryostats are connected by a cold tunnel and share the same vacuum.}
\label{general}
\end{figure*}

\begin{figure}[ht]
\begin{tabular}{c}
\includegraphics[width=80mm]{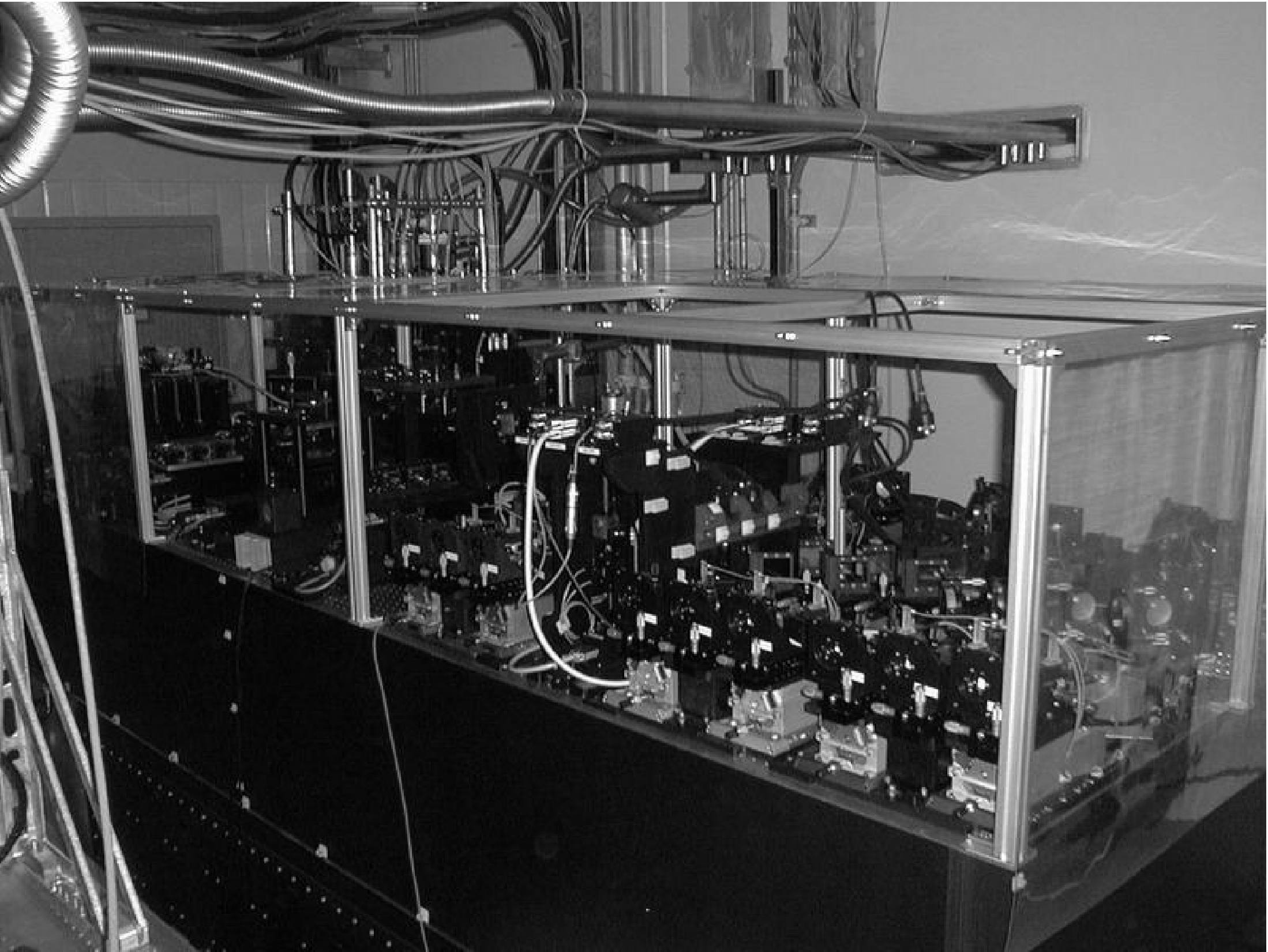}
\end{tabular}
\caption{Picture of the AMBER instrument at the end of the integration at Paranal in March 2004 (by A. Delboulb\'e, LAOG).}
\label{photo}
\end{figure}

Figure~\ref{general} shows the global implementation of AMBER, and Fig.~\ref{photo} shows a picture of the instrument taken at the end of the integration at Paranal (March 2004). The 

 of the instrument is composed of the following modules, filling specific functions:\newline
\indent- SPatial Filters (SPF) to spatially filter the wavefront perturbation and reach high-visibility precision measurements. The functions of this element are also: spectral band selection (J, H, and K), interferometric arm selection, control of the beam size and position, flux optimization, OPD equalization,  polarization control, and combination of the spectral bands.\newline	
\indent- ANamorphosis System (ANS) to compress the beams in one direction without perturbing the pupil location.\newline
\indent- Cooled SPectroGraph (SPG). This element includes: dispersion with different resolutions, thermal noise reduction, pupil configuration, photometric calibration, and spectral filtering.\newline	
\indent- Cooled DETector (DET), which detects the dispersed fringes. \newline 
The auxiliary modules are:\newline
\indent- System to correct the atmospheric transversal dispersion (ADC) in J and H.\newline
\indent- Calibration and Alignment Unit (CAU) necessary to perform the contrast calibration \citep{millour,tatuc}.\newline
\indent- Remote Artificial Sources (RAS) allowing for the alignments, the spectral calibration, and feeding the CAU for the contrast calibration.\newline
\indent- Matrix Calibration System (MCS) scheduled to calibrate the contrast to achieve specific scientific goals.\newline
\indent- A BYPass (BYP) of the SPF to align the warm instrument in the visible (for controlling the pupil, the image location, and the beam separation and height), and to inject light directly to SPG.\newline
We will describe each module in detail in the following paragraphs, from the entrance of AMBER to the detector, starting with the main modules and continuing with the auxiliary ones.

\section{The spatial filters}
The three VLTI beams at the entrance of AMBER have a diameter of 18 mm and equal separations of 240 mm (see Fig.~\ref{general}). The three AMBER beams are separated by 70~mm at the fiber entrance. The separations at the entrance are achieved by adjusting the VLTI beam injection optics, allowing us to compensate for the optical path difference with additional path lengths. The chosen configuration ensures perfect symmetry between the interferometric paths and allows for the use of small size optics.

\subsection{Characteristics of the spatial filtering} 
\begin{figure} [h]
\includegraphics[width=42mm]{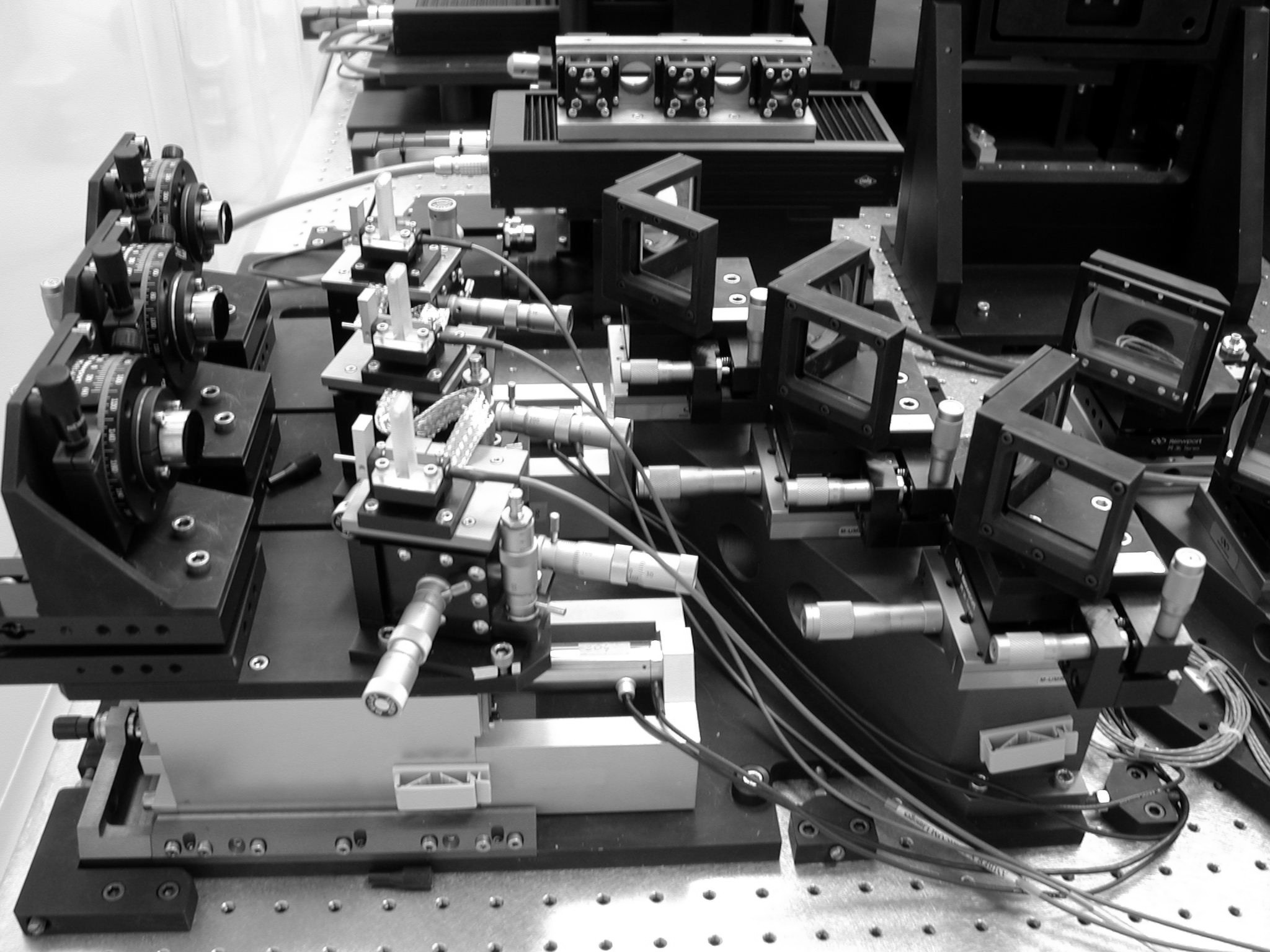}
\includegraphics[width=42mm]{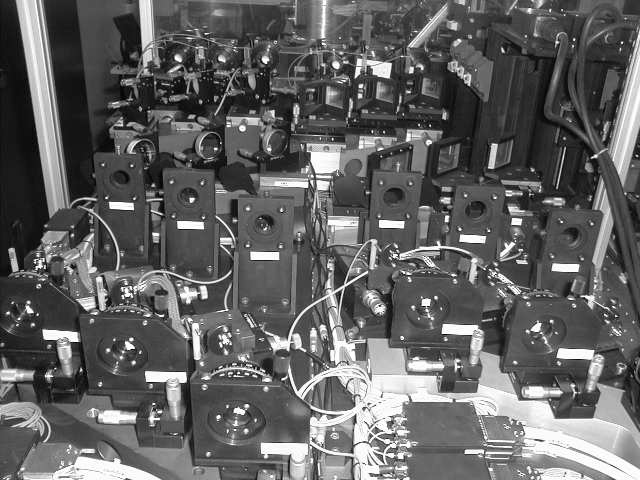}
\caption{Pictures of the K-spatial filter entrance. Left (picture by Y. Bresson, OCA):  the three beams meet the dichroics and the parabolic off-axis mirrors before the injection through the Si birefringent single-mode fibers; Right (picture by A. Delboulb\'e, LAOG): the optical configuration is repeated at the exit of the fibers. Diaphragms control the beam size at the exit of the fibers and shutters select the interferometric arms.}
\label{SF}
\end{figure}

Single mode fibers cannot be efficient over a too large wavelength domain. The full J, H, K range from 1 to 2.4~$\mu$m needs at least two different fibers. The most efficient way is to use one spatial filter by spectral band, avoiding dividing the H-band in two, which would result in the loss of a part of the H-band. The spatial filtering modules SFJ (H, K) (Fig.~\ref{SF}) receive the light from dichroics. Parabolic mirrors inject the light in silicate birefringent single-mode fibers. At the exit of the fibers the same optical system is repeated. 
\subsubsection{The optical fibers}
LAOG provided equalized J-fibers, H-fibers, and K-fibers with a maximal accuracy of $\pm$20 $\mu$m. The characteristics of the presently used fibers are given in Table \ref{fiber}.  
\begin{table} [h]
\caption{Manufacturer and characteristics of the fibers in the K-, H-, and J-bands. NA: numerical aperture, $\lambda_c$: cut-off wavelength, $\oslash$: diameters, Conc.: core concentricity.}
		\begin{tabular}{l l l} \hline\hline
\scriptsize K-band &\scriptsize	Highwave	&\scriptsize NA=0.16; $\lambda_c$=1900nm \\	
\scriptsize  &\scriptsize	Silica	&\scriptsize	Core $\oslash$=9.7$\mu$m; Conc. $<$ 5$\mu$m
\\
\scriptsize &\scriptsize Elliptical core	&\scriptsize Coating $\oslash$=245 $\mu$m\\
\scriptsize &\scriptsize 	&\scriptsize Absolute length=1.30 m $\pm$0.01 m\\
\scriptsize &\scriptsize 	&\scriptsize Length difference after polishing : $\leq11\mu$m \\\hline

\scriptsize H-band &\scriptsize	Fujikura	&\scriptsize NA=0.15; $\lambda_c$=1150nm \\	
\scriptsize  &\scriptsize	Silica	&\scriptsize	Mode $\oslash$=5.5$\mu$m$@$1300nm; Conc. 0.2$\mu$m
\\
\scriptsize &\scriptsize Panda core	&\scriptsize Coating $\oslash$=245 $\mu$m\\
\scriptsize &\scriptsize 	&\scriptsize Absolute length=1.30 m $\pm$0.01 m\\
\scriptsize &\scriptsize 	&\scriptsize Length difference after polishing : $\leq12\mu$m \\\hline

\scriptsize J-band &\scriptsize	Fibercore	&\scriptsize NA=0.14; $\lambda_c$=944nm \\	
\scriptsize  &\scriptsize	Silica	&\scriptsize	Mode $\oslash$=6.3$\mu$m$@$1060nm; Conc. 0.28$\mu$m
\\
\scriptsize &\scriptsize Bow tie core	&\scriptsize Coating $\oslash$=245 $\mu$m\\
\scriptsize &\scriptsize 	&\scriptsize Absolute length=1.30 m $\pm$0.01 m\\
\scriptsize &\scriptsize 	&\scriptsize Length difference after polishing : $\leq20\mu$m\\\hline
\end{tabular}
\label{fiber}
\end{table}

The specifications on spatial filtering are driven by the quality of the optical fibers. The fiber length of about 1.30~m ensures a good transmission while maintaining a 10$^{-3}$ attenuation of the high order propagation modes \citep{malbet}.
The polarization-maintaining is achieved by fibers with an elliptical core causing the so-called "form birefringence", or by strong birefringence caused by two stress members applied on opposite sides of the core (bow tie- and panda-type).
\begin{figure}[ht]
\begin{center}
\includegraphics[width=87mm]{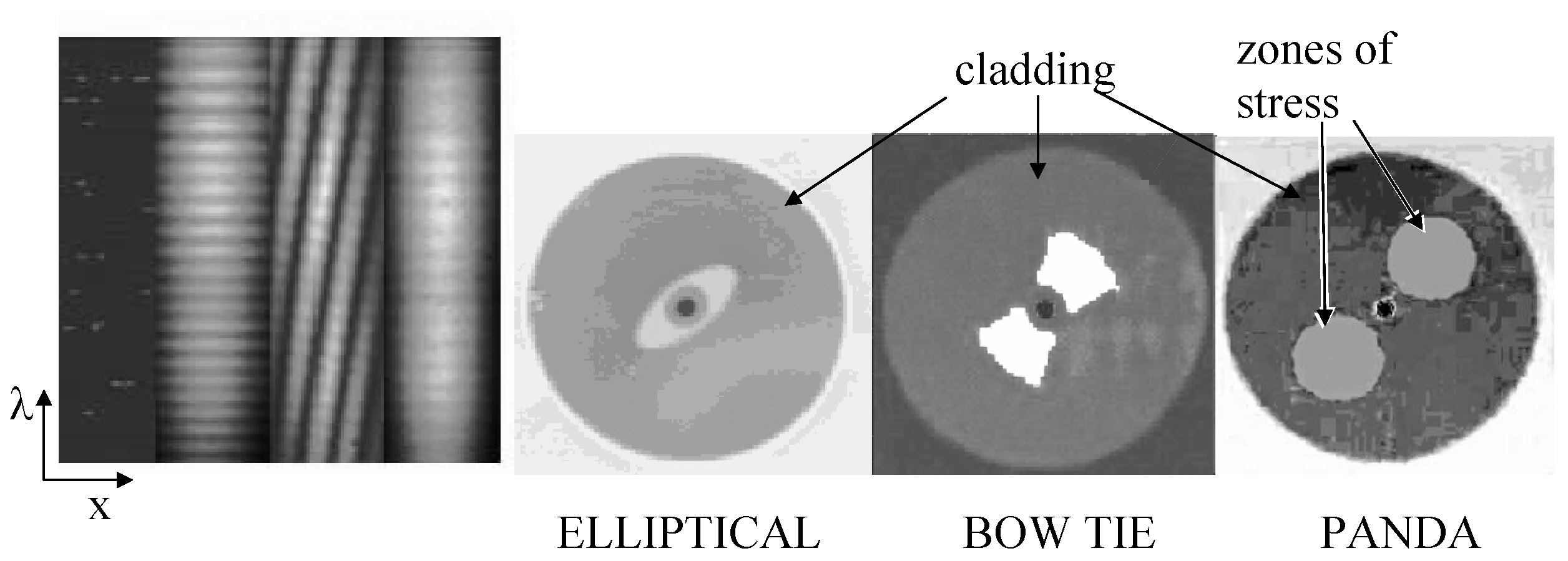}
\end{center}
\caption{Left: illustration of photometric images (extreme) and interferometric image (center) produced with the artificial source (H-band, $\Delta\lambda$=150 nm). The zero OPD is not exact. Note the intensity modulation generated by the fibers. Right: cross-sections of polarization maintaining fibres: elliptical core, bow tie, and panda fibers (http://www.highwave-tech.com/; http://www.fibercore.com/).}
\label{chauss}
\end{figure}
The fibers create an intensity modulation in some of the images recorded by AMBER. Fig.~\ref{chauss} shows this modulation for optical fibers used previously to those of Table \ref{fiber}. This effect, fainter now, is present in the science as well as in the calibrator source and can then be reduced below the specifications in the calibration process. However, it depends on the fiber temperature and on the injection conditions (conditioned by the Strehl ratio). As far as the highest accuracy goals are concerned (especially that of reaching very high accuracy differential phase measurements), a fast correct calibration of this effect is necessary. 

\subsubsection{Fiber injection}
\begin{figure}[ht]
\begin{center}
\includegraphics[width=80mm]{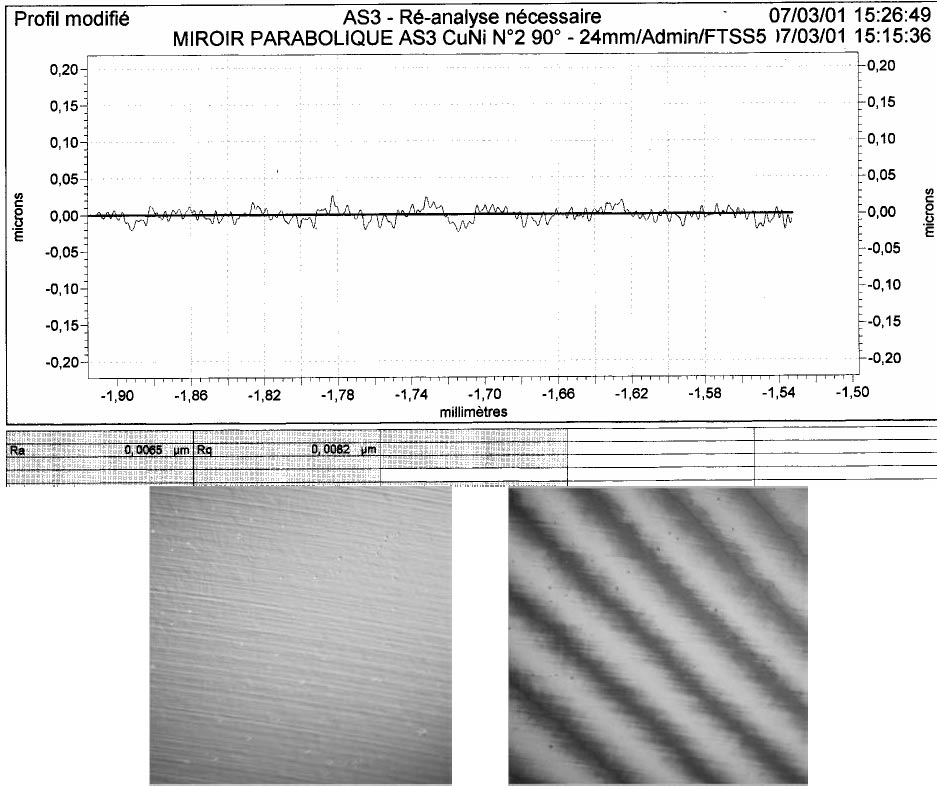}
\end{center}
\caption{Above is the surface profile as measured with the micro-sensor of SAVIMEX on a test element: maximum roughness of 60 nm PTV and RMS roughness of Ra 6.5 nm and Rq 8.2 nm; Below is the picture of the mirror allowing the manufacturing of the three injection optics in the SFJ as observed through the microscope (field of 400x300 µm). The PTV roughness is about 1/4th to 1/8th of fringe, i.e., 80 to 40 nm. The RMS roughness is about 6 nm. The surface optical quality measured with the collimating lunette is below 30 nm, i.e., $\lambda$/20 PTV $@$ 633 nm on an 18-mm diameter disk.}
\label{diam}
\end{figure}

The parabolic mirrors, designed by the French company SAVIMEX (Grasse), are metallic off-axis diamond-turned mirrors. They were controlled by SAVIMEX using a procedure developed in collaboration with OCA using a microscope and a micro-sensor for the roughness, and a collimating lunette for the surface control (Fig.~\ref{diam}). The focal length F is related to the fiber numerical aperture NA and the beam diameter D. The best coupling efficiency was given by Zemax for $NA.F/D=0.46$, taking the telescope obstruction into account. 

\subsubsection{OPD control}
This system was designed by OCA, and manufactured and tested at the Technical Division of the Institut National des Sciences de l'Univers (DT/INSU, Paris).
The static OPD in K is controlled with a 4 $\mu$m accuracy within a few mm range. 
While the static OPD is adjusted for a given wavelength in each band, there exists an OPD drift between wavelengths. AMBER needs to correct this differential OPD in J- and K-bands respective to the H-band (ESO fringe sensor unit functioning) during the observations. This chromatic OPD due to the atmospheric refraction is introduced during the telescope pointing. It is given by: $B\sin z (n(\lambda_{1})-n(\lambda_{2}))$, where $n(\lambda)$ is the refractive index of the atmosphere, $B$ is the baseline, and $z$ the zenithal angle. Considering the extreme case for which $B~=~200$~m and $z = 60^\circ$, the necessary OPD range is about 30$~\mu$m in K and 70$~\mu$m in J. Such adjustments are achieved through drifts of the entire AMBER K, H, or J spatial filter entrance parts. They are performed every few minutes.

Nevertheless, such translations cannot compensate for any chromatic OPD gap present inside each spectral band ($\delta\lambda$ equal to 33 nm in K, 32 nm in H, and 24 nm in J). This chromatic OPD gap is introduced when a difference in the glass thickness or in the fiber length between two interferometric arms exists. Limiting this relative thickness to 0.5 mm, the contrast factor is thus ensured of being less than 0.99.
The relative fiber lengths in all the spatial filters were controlled by LAOG (Grenoble). It was shown that at the minimal resolution of AMBER, the contrast degradation factor due to differential dispersion is better than 0.99 inside each spectral band \citep{robc}.

\subsubsection{Rapid OPD variations} 
Instabilities of the VLTI beams can generate rapid achromatic and chromatic OPD changes. The first type of problem results in a deviation of the entire fringe pattern sideways. This is compensated by the VLTI itself. The chromatic changes degrade the interference pattern, curving the fringe shape at the timescale of these instabilities. It leads to the presence of a blurred pattern, especially at the sides of each spectral band (even if the OPD is well stabilized at the central wavelength of each spectral band). The main source of such instabilities comes from the differential positioning shifts of the VLTI beams, combined with the travel of light towards wedged dispersive glasses. Consequently, the requirement is to ensure that dynamic chromatic OPD rms values are lower than the uncertainty that comes from the fundamental noise levels. The goal is to reach the (mainly photon) noise corresponding to a 1-minute measurement with a 5-magnitude star (observing with 2 UTs and the AO of the VLTI). In terms of chromatic OPD, between the central wavelength and the wavelength at the side of the considered bandwidth (most demanding case: R=35), this translates into OPD(noise)~=~5.8 e$^{-11}$ m in K. \citet{vanniera,vannierb} performed a complete study that led to the instrument requirement analysis concerning the wedge angle of the prismatic optics, and the surface quality and the relative thickness of the dispersive elements located prior to the fibers. This study included elements such as optical fibers, polarizers, and dichroics, but also those of the auxiliary modules described below in this paper, such as the Neutral Densities (NDN), the Matrix Calibration System (MCS), and the Atmospheric Dispersion Corrector (ADC).

\subsubsection{Dichroics}
The dichroics satisfy the photometric requirements: $\geq$~0.95 in reflection in the highest spectral band and $\geq$~0.90 in transmission in the lowest spectral band. An absorption is present at the end of the H-band (transmission from 86\% at 1800 nm to 80\% at 1850 nm), but it does not affect the global throughput.

\subsubsection{Polarization control}
The following was done to minimize the polarizations' effects: \newline
\indent - Use the same number of reflections/transmissions between the 2 interferometer beams.\newline 
\indent - Require identical coatings for the optics with the same functions between 2 beams: same substrate, same structure for the different layers, simultaneous manufacturings.\newline
\indent - Use polarization-maintaining fibers.\newline
\indent - Select one polarization direction at the fiber entrance.\newline
\indent - Control the incident angle on reflecting optics with an accuracy better than a fraction of a degree.\newline

Prior to the spatial filters, the polarizers select one polarization direction to get rid of the cross-talk inside the fibers and of the phase difference between beams (variable differences during the telescope pointing). The selected direction is that which is not affected by the multiple reflections inside the instrument (perpendicular direction to the beam propagation plane). 

Each polarizer is associated with one blade. The orientation of the two elements is mechanically controlled to ensure the direction of the light beam relative to the optical axis. The air blade located between the two prism constituents of each polarizer must be parallel for the optical system to respect the chromatic dynamic OPD specifications (see Sect. 3.1.4). 

The relative polarization control also has an impact on the optical coating quality, in particular for the dichroics elements. All the optics with the same functions are simultaneously coated by the manufacturer, in particular the dichroics and the injection parabola located between the polarizers and the fiber entrance, to reach a flux difference of a few \% after each reflection or transmission and a minimal phase difference generated by the different layer thicknesses.

At the spatial filter exits, an error on the dichroics layer thickness could generate a contrast degradation. The number of elements being small (1 reflection/arm in K, 1 reflection/arm and 1 transmission/arm in H, and 2 transmissions/arm in J), the overall effect is almost null. From \citet{puech}: a typical 2\% of error on the layer thickness generates less than 1\% contrast loss in K. The neutral axes at the fiber entrance are controlled with a $\pm$3$^\circ$ accuracy to compensate for the polarization direction rotations generated by the residual manufacturing differences between dichroics and by the incident angle differences. This ensures the maximum coherent energy inside the fibers. At the fiber exit, the differential polarization between beams before combining is controlled to within a few degrees. 

\subsubsection{Acoustic perturbations}
Acoustic perturbations (due to step-by-step motors for instance) can modify the behavior of optical fibers that are sensitive to pressure variations.
It can be shown that a typical talk produces a 60 dB acoustic intensity, which implies a phase instability of 10$^{-7}$ rad, far below the specification (Perraut, internal report).
To avoid disturbing other VLTI equipment in the interferometric laboratory, each instrument does not generate acoustic noise in excess of 40 dB at 2 m in all the directions.

\section{From the spatial filters to the detector}
The general parameters of the modules from the spatial filters to the detector are given: beam configuration, pupil diameter and separation, spectral resolution, and signal sampling.
The modules are then described in details.
 
\subsection{General parameters of the system modules}
\begin{figure}[ht]
\begin{center}
\includegraphics[width=75mm]{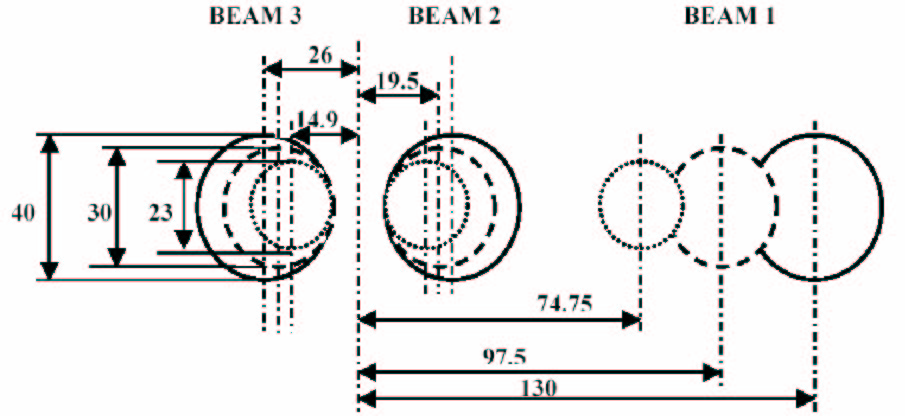}
\end{center}
\caption{AMBER pupil configuration in J, H, and K with no anamorphosis.}
\label{pupil}
\end{figure}

\underline{Beam configuration}: AMBER is a "dispersed fringes" instrument operating in the image plane. The three-pupil configuration (Fig.~\ref{pupil}) is a non redundant line set-up. 
The maximal baseline is larger than the minimal one by a factor three. This pupil configuration produces three systems of fringes corresponding to the following baseline: $B_{m}$, $2B_{m}$, $B_{M} = 3B_{m}$ (the indices $m$ and $M$ mean minimum and maximum, respectively).
\underline{Pupil diameters and spectral resolution}:
The pupil size depends on the spectral resolution and on the number of grooves per mm of the grating. The central wavelength ($\lambda_{0}$ = 2.2 $\mu$m in the K-band) involves a limitation of the number of lines per millimeter for the spectrograph grating (about 500 lines/mm). To optimize the recorded flux, the spectral channel is a bit undersampled ($\lambda_{0}$/D on one detector pixel). To obtain a spectral resolution of 10\,000, the pupil diameter in the K-band is $D$ = 40 mm. The fringe sampling is the same for all the spectral bands. This implies smaller pupil diameters in the other spectral bands (30 mm for the H-band and 23 mm for the J-band). The instrument pupil is set by a cold stop inside the spectrograph. This pupil plane is combined with the neutral point of the cylindrical optics, roughly superimposed to the pupil masks located after the collimating parabola at the exit of the spatial filters. The precision of this conjugation has a negligible impact on the performance of this single mode instrument with a field of view limited to an Airy disk.

\begin{figure} [h]
\begin{center}
\includegraphics[width=60mm]{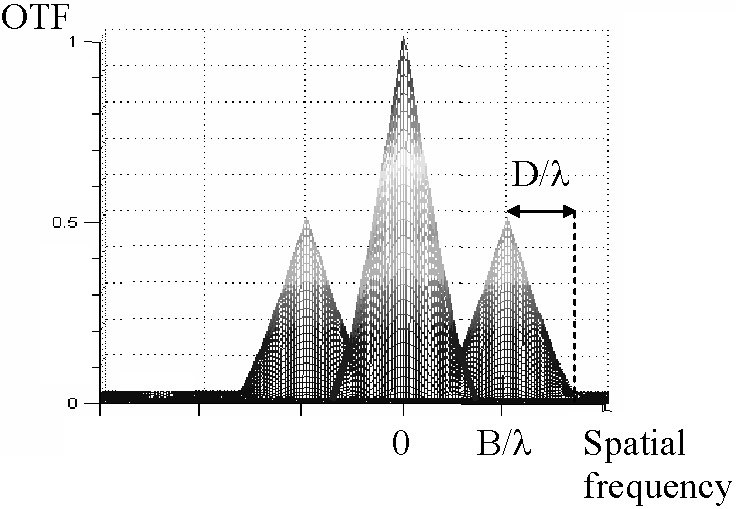}
\end{center}
\caption{Illustration: simulated optical transfer function (OTF) in which the coherent and incoherent energy peaks are not completely separated. The pupil separation is chosen such that the fringe peak center is not affected by the incoherent peak.}
\label{otf}
\end{figure}

\underline{Pupil separation}:
In the optical transfer function (OTF), the fringes with the lowest frequency produce a coherent energy peak close to the incoherent energy peak (Fig.~\ref{otf}). To avoid the center of this fringe peak to be affected by the central single pupil peak, the minimal baseline $B_m$ is: $B_{m}>D(1+\lambda_{M}/\lambda_{m})$. For a full band observation in the K-band ($\lambda_{m}$ = 2.0 $\mu$m and $\lambda_{M}$ = 2.4 $\mu$m), $B_{m}$ must be greater than 1.2$D$.
For this reason, the distances between the three pupils are 1.3$D$, 2.6$D$, and 3.9$D$.

\underline{Signal sampling: anamorphosis factor and camera fo-}
\underline{cal length}: Considering data reduction requirements, it is necessary to sample the fringes produced by the combination of the furthest beams by about four pixels on the detector. Each spectral element ($\lambda_{0}/D$) is analyzed by one pixel. The magnification of the beams between the spatial direction and the spectral direction must be different. This anamorphosis factor $a$ is given by: $a = 4 B_{M}/D = 15.6$. The size $p$ of the detector pixel, equal to 18.5 $\mu$m, is linked to the camera focal length $f_{c}$ by the relation $4~p~=~a~f_{c}~(\lambda_{0}/B_{M})$. The $f_{c}$ parameter is then deduced from:  $f_{c} = p /( \lambda_{0}/D) \approx  350$~mm.

\subsection{Anamorphoser system (ANS)} 
At the exit of the spatial filters, the beams enter the cylindrical optics anamorphoser (ANS) before entering the cold spectrograph (SPG) through a periscope and a focalizing optical component adjusting both axes and image positions at the interface between the warm optics and SPG.
The anamorphoser is a Chretien hypergonar made of an afocal system of two cylindrical mirrors inserted in a parallel beam section. The anamorphosis factor, 15.6, is the ratio between the focal length of the two mirrors. The anamorphosis direction is perpendicular to the baseline.
Such an afocal system contains a neutral point, located nearby after the focal plane of the smallest (2$^{nd}$) cylindrical mirror. By putting the conjugate of the spectrograph cold stop at this neutral point, we avoid having the ANS introduce a difference between the longitudinal and transverse pupil positions, minimizing the aberrations and improving the cold stop baffling.
The difficulty lay in the manufacturing of the 1$^{st}$ optics of the ANS shared by the 3 interferometric beams: a conic 220x50 mm rectangle with a 2-m curvature radius in its length direction. The PTV optical quality of 633 nm/5 was tested on a specific optical bench at OCA.

\subsection{The spectrograph (SPG)}
The cold spectrograph SPG includes the following functions:\newline
\indent - Filtering of thermal radiation at the input image plane. \newline
\indent - Formation of a parallel beam.\newline
\indent - Accurate spatial filtering of the pupils.\newline
\indent - Separation of interferometric beams from photometric beams.\newline
\indent - Spectral analysis at three resolving power values.\newline
\indent - Formation of images on the detector plane.\newline
The list of the SPG is given in Table \ref{SPG}, with the product tree definitions and functions.
\begin{table} [h]
			\caption{List of AMBER spectrograph modules with acronym definition and functions.}
		\begin{tabular}{l l } \hline\hline
\scriptsize \textbf{Optical elements} &\scriptsize	\textbf{Functions}	\\	\hline
\scriptsize SPG-INW &\scriptsize Input slit.\\
\scriptsize Input Wheel &\scriptsize Image cold stop.\\ \hline
\scriptsize SPG-CSY &\scriptsize Spectrograph optics cooling.\\
\scriptsize Cooling System &\scriptsize Vacuum, temperature, and nitrogen\\
&\scriptsize level control.\\ 
&\scriptsize Thermal flux reduction.\\ \hline
\scriptsize SPG-ISD &\scriptsize Cold stop for the spatial \\
\scriptsize Imaging and Stopping  &\scriptsize filter mode (thermal flux reduction).\\
\scriptsize Device&\scriptsize Technical operations (wide diaphragm).\\ 
&\scriptsize Calibration (dark).\\ 
&\scriptsize Beam collimation.\\ \hline
\scriptsize SPG-PMW &\scriptsize Thermal flux reduction.\\
\scriptsize Pupils Masks Wheel &\scriptsize Beam size definition.\\ \hline
\scriptsize SPG-IPS &\scriptsize Thermal flux reduction.\\
\scriptsize Interferometric Photometric &\scriptsize Beam splitting.\\ 
\scriptsize Splitter &\scriptsize Beams deviation.\\ 
&\scriptsize Photometric images deflection on DET.\\ \hline
\scriptsize SPG-DIU &\scriptsize Spectral dispersion.\\
\scriptsize Dispersion Unit &\scriptsize Spectral wavelength selection.\\
&\scriptsize Spectral resolution selection.\\ 
&\scriptsize Spectral modulation.\\ \hline
\scriptsize SPG-CHA &\scriptsize Beam combination.\\
\scriptsize Camera&\scriptsize Beam focalization.\\
&\scriptsize Sampling.\\ \hline
		\end{tabular}
			\label{SPG}
\end{table}

\begin{figure} [ht]
\begin{center}
\includegraphics[width=75mm]{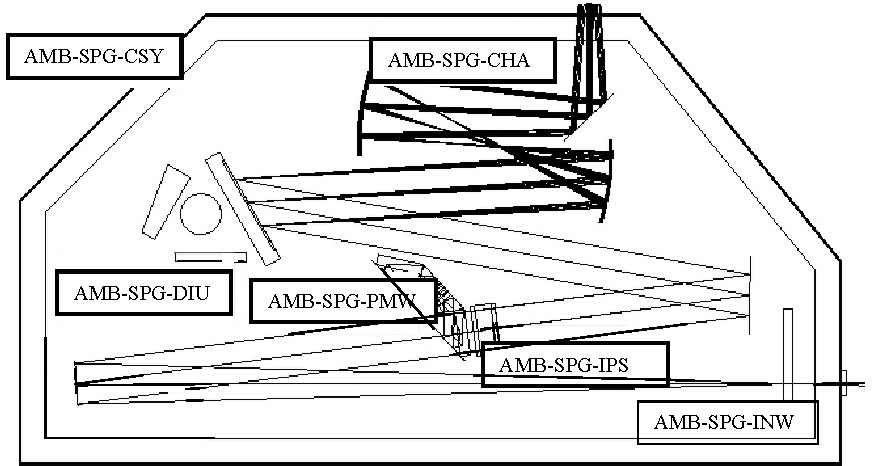}
\end{center}
\caption{Optical design of the SPG. For the acronym definition see Table \ref{SPG}. }
\label{spg}
\end{figure}

The optics and the optical bench are contained in a vacuum tight cryostat
that allows the cooling of all functions to the working temperature of
about 77 K by means of liquid nitrogen at atmospheric pressure.  The
spectrograph cryostat does not lodge the detector that stays in a second
cryostat; the two cryostats are mechanically coupled, they
share the same vacuum and work at the same temperature, but have two
independent cooling systems. The coupling is achieved by a flexible bellow
and an interface structure that allows for a certain degree of angular
and linear adjustment to align the detector itself to the spectrograph
optics.
The optical design of SPG (Fig.~\ref{spg}) follows the general pattern of the
grating spectrograph, with the necessary modifications dictated by the
optical and mechanical accuracy requested by interferometry. A more
detailed description of SPG is in \citet{lisi}.

\subsubsection{SPG optical configuration}
The OPM-OSI module (Fig.~\ref{general}), located at the side of the SPG cryostat, injects the
beam into the spectrograph through a CaF$_2$ window. The beam is spatially
filtered by the slit, mounted on a wheel that also carries
several components useful to the calibration and alignment operations. \newline
A collimating mirror produces the parallel beam. The collimating mirror is a symmetric paraboloidal mirror (with respect to the focal point) of the focalizing mirror; this configuration has the purpose of minimizing the optical aberrations. It has a focal length of 700 mm and an aperture of F/3.   \newline
The wheel SPG-PMW carries the pupil masks, located on the plane of the pupil image, which
help to minimize the stray thermal background. The optical component
SPG-IPS is a beam splitter that steers part of the flux coming from each
telescope from the parallel beam for photometric calibration, leaving the
largest fraction of the flux for the interferences. The beams that
carry the photometric information run along the same optical path as the
interferometric beams, but they are suitably directed to form separate
images on the detector. \newline
For the purpose of spectral analysis, a rotating device
(SPG-DIU) allows for the choice among three dispersing components: two
gratings (497 g/mm and 75 g/mm for respective resolutions of 10\,000 and 1\,500), and a prism. 
The support of the gratings and the prism is motorized to allow us to select the spectral range and resolution. The angular accuracy of this support is about 3 pixels of the detector ($\approx$ 30''), which implies a spectral calibration procedure. \newline
The final optical function before the detector is the camera SPG-CHA, designed around
three mirrors. It is composed of two aspherical mirrors with a total focal length of 350~mm (see Sect. 4.1.4) and an aperture of about F/2. A plane mirror steers the beams coming from the camera unit to send it to the DET cryostat. A spectral filter is inserted in the J pupil mask to eliminate the background coming from the K-band, while observing with the second spectral order of J. 

\subsubsection{Opto-mechanics and performance}
The requirements for the optical design reflect the need to mount all the
optics inside a cryostat, where the accessibility for alignment is reduced
and the displacements of components after the cooling are very large.
The tolerance analysis shows a fringe contrast degradation factor of
about 95$\%$ under the following constraints: a total
positioning error of the optical elements of 0.1 mm, an
angular positioning error (tilt) of 2.3 mrad, and a quality of optical surfaces better than 5 fringes in focusing and 1 fringe in irregularity. This performance depends only on the use of the
detector position along the optical axis as a compensator, with a total
displacement of less than 2.3~mm with respect to the nominal position.
The surface quality (micro-roughness) of the metallic mirrors has an
impact on the efficiency. The machining of the aspherical and plane
mirrors allowed us to produce the respective roughnesses of less than 10 nm and 5 nm rms, to ensure that the loss of light be less than 1.5$\%$ on each mirror. One feature of the opto-mechanical design is keeping the optics aligned at room
temperature and at liquid nitrogen temperature without adjustments.
Tests of the SPG optics either at room or at operative temperature showed no significant differences of performance between the two sets of
measurements, confirming the design concept.
\subsubsection{Vacuum and cryogenic system} 
The whole system is lodged inside a vacuum-tight case made out of welded
steel plates with suitable reinforcing ribs.  The liquid nitrogen vessel is
a box-like structure (worked out of a single aluminum block completed by a
welded cover), whose bottom plate is the cold optical bench.  The external
case supports the cold bench by means of an hexapod (composed of six fine
steel beams), dimensioned to allow the SPG system to be placed on a side
without undergoing permanent deformation. All the optics are enclosed in the radiation shield, in tight
thermal contact with the cold bench. The model of the thermal behavior
shows that all its points are at most two degrees over the cold bench
temperature; this is confirmed by measurements. The SPG optical system includes a total of three moving wheels, the
aperture wheel, the pupil mask wheel, and the grating wheel. To simplify
the engineering, these three functions are implemented by cryogenic
motors (Berger Lahr 5-phase), modified according to the
ESO experience. Positioning of the associated wheels is performed by a
worm-wheel gear that is substantially irreversible and acts as a stop to
the force exerted by the spiral spring used to overcome the backlash. The
temperature of the three motors is constantly monitored by the control
system using local Pt100 sensors. The nitrogen vessel can store up to 19 dm$^3$ of liquid N$_2$, while the mass
of aluminum to be cooled is about 24 kg.  The design total thermal load is of
the order of 25 W.  A turbo-molecular pump establishes the operative
vacuum and in normal operation, a small quantity of active charcoal keeps
the internal pressure at the level of 2.10$^{-5}$ mbar for several
months. The liquid nitrogen supply lasts for about 30 hours.
\subsection{The detector (DET)}
\subsubsection{Hardware overview}
\begin{figure}  [ht]
\includegraphics[width=70mm]{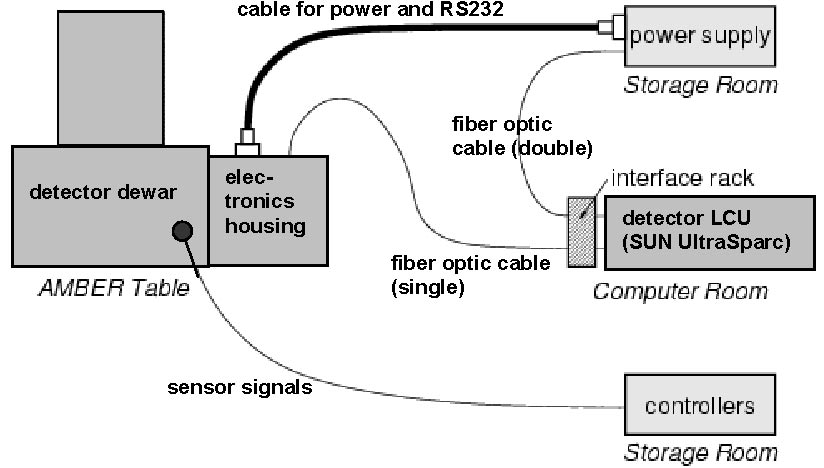}
\caption{Physical overview of the AMBER detector hardware}
\label{det1}
\end{figure}

The detector is located in a dewar that is cooled down to 77 K with liquid nitrogen. The detector electronics housing is directly attached to the dewar to avoid electronic interference resulting from long signal paths (Fig.~\ref{det1}). It is connected to the sensor by two short cables. In addition to this, it complies with challenging constraints concerning heat dissipation, interference, and electromagnetic compatibility, for example.
The power supply is not installed close to the detector electronics housing. A distance of a maximum of 15 m is allowed between the electronics housing and the power supply. The connection is made by a single power cable. This cable also contains the galvanically isolated RS232 serial line for controlling the electronics. The power supply rack is installed in the instrument control cabinet in the storage room. The digital image data is transmitted to the detector LCU via a fiber optics cable.

\subsubsection{Functional overview}
The detector electronics consists of the following modules:\newline
1.	An infrared detector (HAWAII-1 focal plane array from Rockwell).\newline
2.	A sequencer generates clock patterns necessary for reading and sampling the sensor. It can be configured by data sent through the galvanically isolated serial line (RS232). It also generates a header containing information about image format, readout mode, etc.\newline
3.	A clock driver boosts the digital signals from the sequencer.\newline
4.	A video amplifier supplies all necessary bias voltages and prepares the analog signal from the IR sensor for sampling.\newline
5.	An analog to digital converter ADC samples the analog signal. On the ADC board there is also digital logic for averaging several samples (subpixel sampling). A fiber optical transmitter on the same board feeds the image data into a fiber optics cable connected to the detector LCU.\newline
6.	A power supply.

\subsubsection{Detector characteristics}
Detector type:	Rockwell HAWAII-1 FPA (focal plane array, one quadrant in use)\newline
Detector size:	512 x 512 pixels\newline
Pixel size:	18.5 $\mu$m x 18.5 $\mu$m\newline
Operating temperature:	77 K\newline
Quantum efficiency:	$>$ 50 \% for 1 - 2.4 $\mu$m wavelength\newline
Other properties of the detector chip were measured and are listed below:\newline
\indent- Detector number:	\#159\newline
\indent- Full well capacity: 63670 e- \newline
\indent- 1\% nonlinearity: 26289 e-\newline
\indent- Conversion factor: 4.70 $\mu$V/e-\newline
\indent- Readout noise (CDS $@$500 kHz): 11.6 e-\newline
\indent- Number of bad pixels: 1489\newline
\indent- Clusters of bad pixels ($\geq$4): $\approx$10

\subsubsection{Image on the detector}
\begin{figure}[ht]
\begin{center}
\includegraphics[width=70mm]{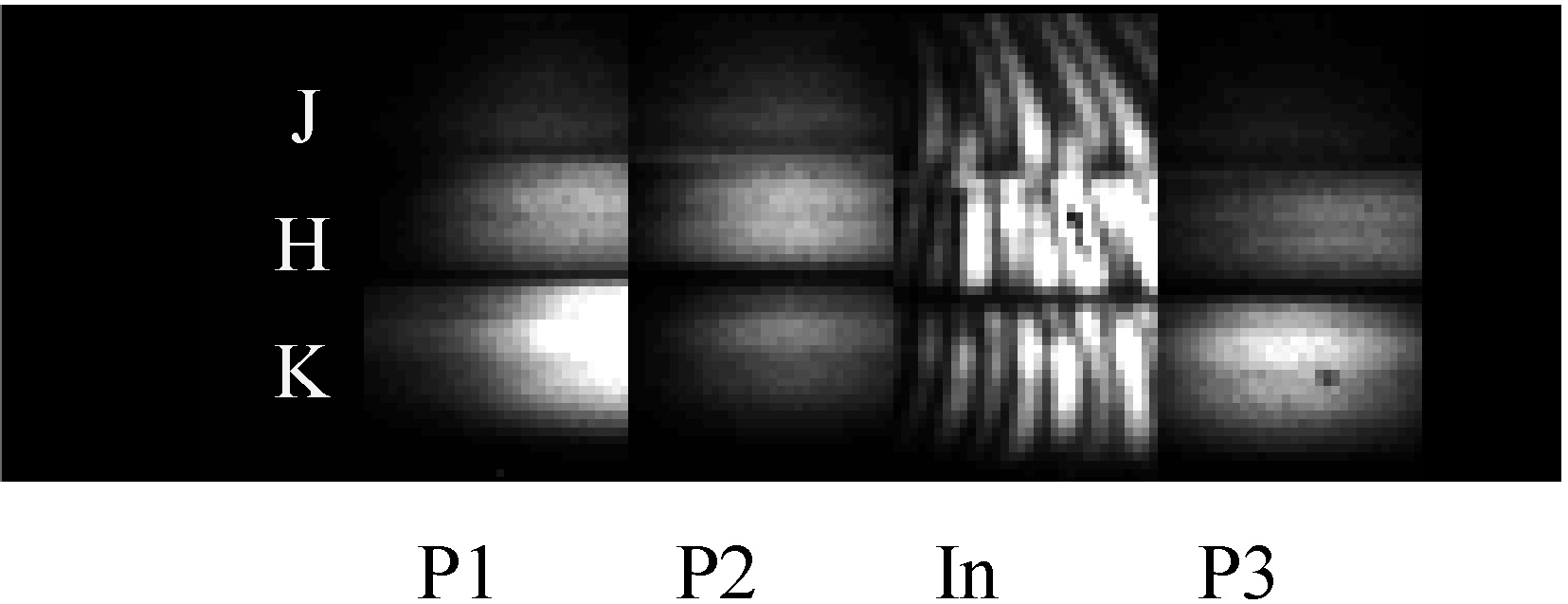}
\end{center}
\caption{Image with the CAU light: dark current, photometric beams (P*), and interferometric beam (In).}
\label{image}
\end{figure}

Figure \ref{image} shows one image recorded on the detector with the CAU light. From left to right  are visualized the dark current, two photometric beams, the interferometric beam, and the 3$^{rd}$ photometric beam. The flux in the J-band was not optimized.

\section{Auxiliary modules}
\subsection{Remote artificial sources (RAS) and calibration and alignment unit (CAU)} 
\begin{figure} [ht]
\begin{center}
\includegraphics[width=84mm]{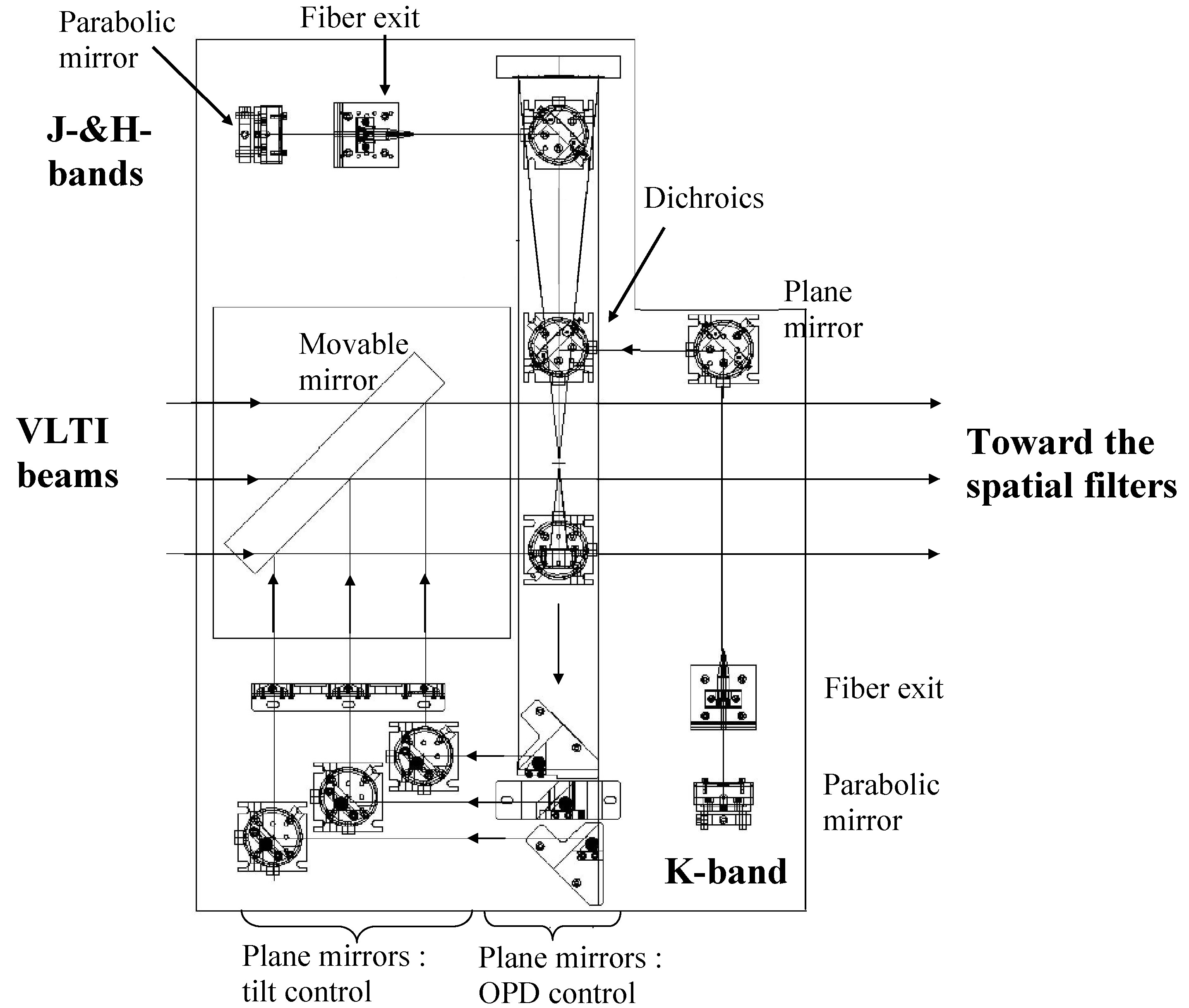}
\end{center}
\caption{Calibration and Alignment Unit (CAU)}
\label{cau}
\end{figure}

Artificial sources are provided in the module RAS for alignments in the visible, flux and OPD control, contrast, and spectral calibrations. These sources are: one laser diode and one halogen lamp allowing alignments and calibration of the matrix of the "pixel to visibility" linear relation (P2VM). The halogen source feeds two different single-mode fibers, one dedicated to the K-band, the other to the J- and H-bands, to transport the light up to the CAU, which can simulate the VLTI in the integration and test phase (Fig.~\ref{cau}). The use of the same fiber in J and H is a compromise solution allowing us to save space and money without losing too much (a few tens of a \%) injected light in both bands. The exits of the two fibers coming from the RAS provide almost point-like sources in J, H, and K. 

The retained optical configuration of the CAU chosen on the basis of generating an achromatic path length uses a wavefront division configuration. The spectral beams exiting the RAS fibers are collimated and recombined in a global wavefront. The latter is magnified and divided in three parts via a set of plane mirrors to be injected inside AMBER via a movable 45$^\circ$ mirror. The equivalent K-magnitude of the CAU was estimated to be -1.4 for the medium beam of AMBER and 0 for the extreme beams (different because of the Gaussian distribution of the global wavefront). The instrumental contrast generated with the CAU light is not 100\%. Given that the cores of the fibers have a finite dimension, the CAU does not provide perfect unresolved sources. Nevertheless, this contrast (from 0.75 to 0.87 depending on the spectral resolution and baseline) is taken into account in the P2VM procedure and does not affect the instrumental contrast of AMBER illuminated by the VLTI. \newline
The CAU light is also used to perform a calibration of the medium spectral resolution. A blade can be inserted at the spectrograph entrance generating Perot-Fabry-like interference fringes with a 30\% contrast and a periodicity of about 0.05 $\mu$m.
In low spectral resolution, we use the spectra as they are defined by the three J, H, and K dichroics transmission curves and calibrated with spectroscopic reference stars and lamps. We successively observed the spectrum of each individual spatial filter by closing the shutters of the two others, which yields a subpixel calibration with an accuracy of about 0.01 $\mu$m, enough for this low resolution mode. 
The highest spectral calibration mode can only be calibrated using spectroscopic calibrators and/or telluric lines like any other high resolution infrared spectrograph.

\subsection{Calibration system} 
To calibrate the P2VM it is necessary to introduce a controlled phase delay (between $60^\circ$ and $120^\circ$) between the interferometric arms. In the present state of the instrument, the piezoelectrics used for the chromatic OPD control are accurate enough (a few nanometers) to perform this phase delay.
Nevertheless, a requirement of $10^{-4}$ rad on the repeatability of the phase value was initially defined to reach some differential interferometric goals. To achieve this performance, a specific set-up (MCS) was designed and could be used in a near future if necessary. It consists of couples of slighty inclined glass blades placed in each beam path, the second one having a tilt opposite to the first one (leading to a V glass shape in each beam path). This system allows for very good tolerance on the exact absolute thickness of the blades, as their thickness variations will be compensated for through a slight inclination control during the optics mounting. This type of configuration enables a repeatable phase delay even if the system undergoes some inclination during its positioning.

\subsection{Atmospheric dispersion corrector (ADC)} 
\begin{figure}[ht]
\includegraphics[height=11.5mm]{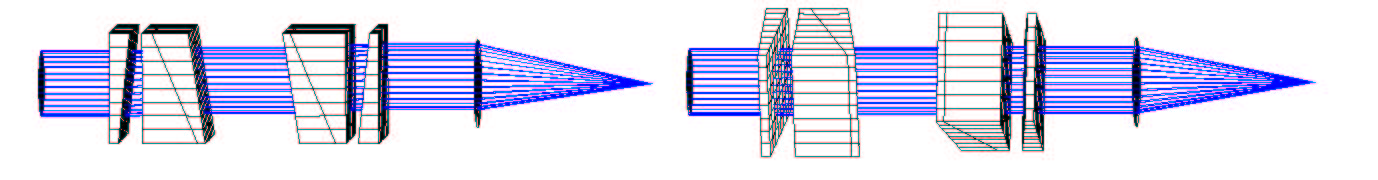}
\caption{Relative axial rotation of one ADC system relative to the other one at z=0$^\circ$ and at z=60$^\circ$.}
\label{adc}
\end{figure}

\begin{figure}[ht]
\begin{center}
\includegraphics[width=70mm]{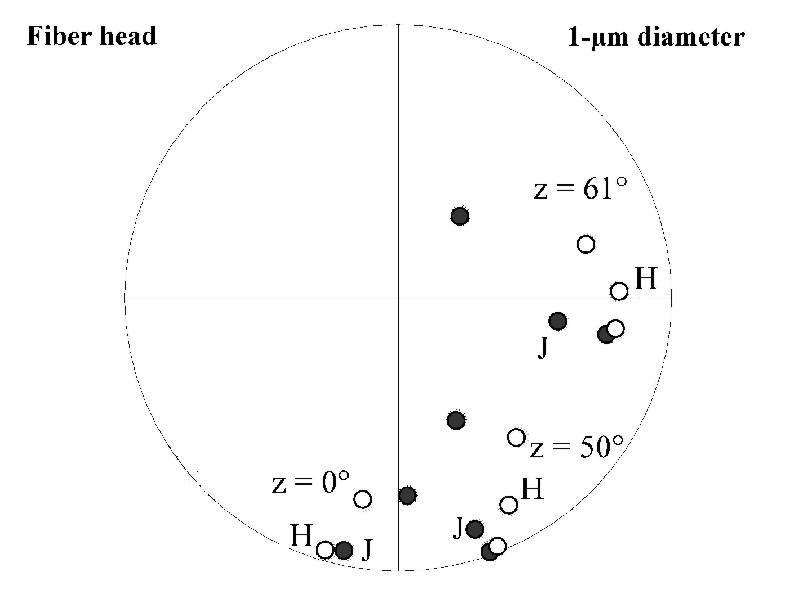}
\end{center}
\caption{Combined optical and positioning errors of the ADC. The spots represent the impacts of the beam for 3 wavelengths inside each band, J (black spots) and H (white spots), for 3 values of the zenithal angle $z$, and taking into account the following errors: prism angles manufacturing, error on the prism gluing, sensitivity of the axial rotation of the two glued prisms respective to the first one, tip/tilt of this rotation axis, tip/tilt of the 3-prisms assembly, sensitivity of the 360$^\circ$-rotation of the two ADC prism system, and inclination induced during this rotation. The worse expected coupling degradation factor is 0.85 in J and H.}
\label{adc3}
\end{figure}

The role of the ADC is to correct the differential transversal dispersion of the atmosphere in J and H. The specific constraints of AMBER require a different conception than the Risley prism generally used. The original system of AMBER is composed of two sets of 3 prisms rotating with respect to each other (Fig.~\ref{adc}) and inserted in each beam prior to the J and H spatial filters. Each system is composed of a first prismatic blade in BK7 and of a doublet of blades in SF14 and F2 glued together.  
During an observation with the UTs of objects located from 0$^\circ$ up to 60$^\circ$ from zenith, the correction of the transversal dispersion is performed by an axial rotation of the two systems. During this rotation, the maximum image shift due to combined optical manufacturing and positioning errors is less than 1 $\mu$m on the entrance fiber heads so as not to affect the coupling efficiency by more than 15\% in J and H (Fig.~\ref{adc3}). This coupling loss is taken into account in the throughput budget.

\subsubsection{Bypass (BYP)} A specific system (BYP) with a commutable mirror was designed to bypass the spatial filters so that preliminary mechanical adjustments and optical alignments in the visible could be performed with technical tools (CCD, lunette) and injected towards the final detector (DET) via the spectrograph. It is also used to ensure that the CAU fibers for the JH- band and the K-band are superimposed.
It appears that the BYP can be used on sky for acquisition of complex sources or technical controls during VLTI troubleshootings. The sampling is 23 mas/sky with UTs and the non-vignetted field about 1 as.

\subsubsection{Neutral densities (NDN)} Two sets of NDN can be inserted in the VLTI beams to avoid the detector saturation. One set is chosen as a function of the target brightness. Two flux attenuations are possible: 10 and 10$^2$.
\section{Optical study and present performance}
The performance of AMBER is given by the signal-to-noise ratio ($SNR$) of the visibility derived from AMBER interferograms. \citet{malbet} showed that the instrument contrast $V_{inst}$ must be 80\% (90\%) and the optical throughput $t_{A}\; C_{eff}$ larger than 2\% (5\%) to reach the magnitude specification (goal) defined for a fringe detection at $SNR$=5. The parameter $t_{A}$ includes both optical transmission and the fiber coupling degradation factor due to misalignments. 
$C_{eff}$ is the fiber coupling efficiency taken to be about 81\%, considering the telescope obstruction.
The other considered parameters in the $SNR$ calculation are:\newline
\indent- $E_0$: flux of a zero-magnitude star at the considered wavelength\newline
\indent- $m$: expected magnitude of the observed object \newline
\indent- $S$: telescope surface area (considering a telescope diameter of 8m for the UTs and 1.8m for the ATs)\newline
\indent- $t_{V}$: VLTI optical throughput (20\% in J, 26\% in H, and 30\% in K)\newline
\indent- $SR$: Strehl Ratio. The $SR$ in K for an on-axis reference source  is equal to 50\% (when the science and reference sources are 1 arc minute away, the Strehl in K is divided by two)\newline
\indent- $\Delta\lambda$: spectral bandwidth\newline
\indent- $\tau$: elementary exposure time ($\tau$ = 10 ms for the high accuracy mode, 50 ms for the high sensitivity mode, and up to 100 s for the long exposure mode)\newline
\indent- $\eta$: detector quantum efficiency (0.6)\newline
\indent- $n$: number of pixels for one visibility measurement (about 16 for 3 telescopes)\newline
The computation of the magnitude assumes the use of a single polarization. 
The integration time $\tau$ depends on AMBER observing modes \citep{malbet}.

In this section, we present the allocations for the contrast and throughout of AMBER with the associated errors budgets. The results of the optical study and the expected instrumental stability are compared with measurements in laboratory. Then the results of the two commissionings at VLTI are discussed.
\subsection{Allocations for the contrast and throughput of AMBER}
\begin{figure}[ht]
\includegraphics[width=90mm]{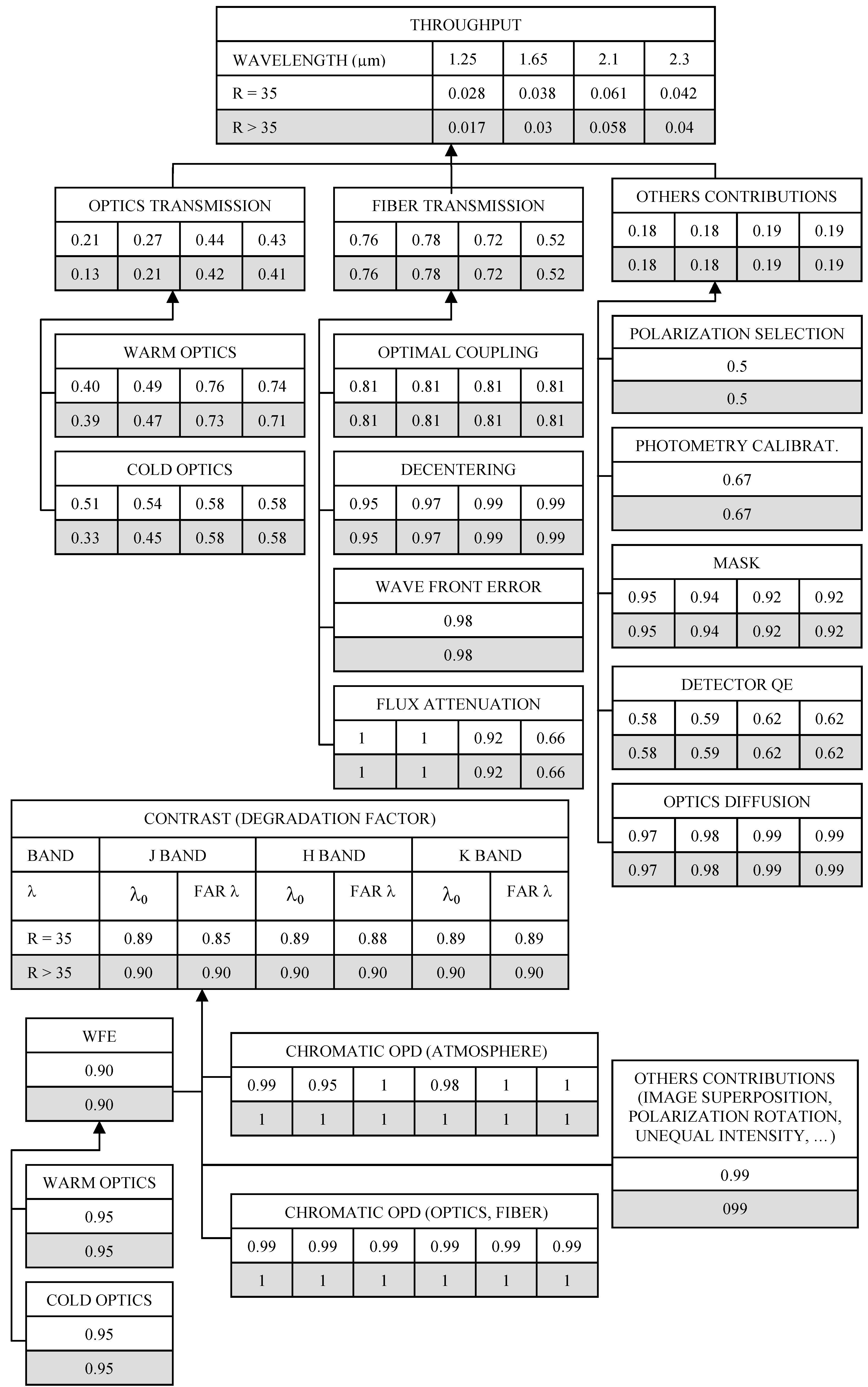}
\caption{Contributions to the throughput and instrumental contrast ($B \sin z=51.2 m$).}
\label{bsinz}
\end{figure}
The diagram of Fig.~\ref{bsinz} summarizes all the identified contributions to the throughput and instrumental contrast ($B~\sin z~=~51.2$~m). 
More details on the allocations are given in Table \ref{detail}. The throughput results are based on conservative values from manufacturer information on coatings.
The next sections give the error budgets on the fiber coupling and the contrast.
\begin{table*} 
\caption{Expected performance of the AMBER design based on the optical study described in this paper.}
\begin{center}
		\begin{tabular}{|l |l |l|l |l|l|l|l|l|l|} \hline
\multicolumn{5}{|l|}{\scriptsize \textbf{Contrast level}}& \multicolumn{5}{l|}{\scriptsize \textbf{80\% to 90\% in K (R = 35) }}\\	\hline
&\multicolumn{4}{|l|}{\scriptsize Static achromatic OPD equality}&&\multicolumn{4}{l|}{\scriptsize $\pm 4 \mu$m} \\ 
\cline{2-5}\cline{7-10}
&\multicolumn{4}{|l|}{\scriptsize Tilt errors}&&\multicolumn{4}{l|}{\scriptsize $\pm$ 3" } \\ 
\cline{2-5}\cline{7-10}
&\multicolumn{4}{|l|}{\scriptsize WF quality}&&\multicolumn{4}{l|}{\scriptsize 140 nm ($\lambda$/16 rms $@$ 2.2 $\mu$m)} \\ 
\cline{2-5}\cline{7-10}
&\multicolumn{4}{|l|}{\scriptsize Differential polarization (after fiber exit)}&&\multicolumn{4}{l|}{\scriptsize $\pm 16^\circ $} \\ 
\cline{2-5}\cline{7-10}
&\multicolumn{4}{|l|}{\scriptsize Differential fiber axis orientation }&&\multicolumn{4}{l|}{\scriptsize $\pm 3^\circ $} \\ 
\cline{2-5}\cline{7-10}
&\multicolumn{4}{|l|}{\scriptsize Bias due to unequal intensity}&&\multicolumn{4}{l|}{\scriptsize 0.9} \\ \hline \hline
\multicolumn{5}{|l|}{\scriptsize \textbf{Minimal optical throughput of AMBER}}& \multicolumn{5}{l|}{\scriptsize \textbf{$>$ 2\% in K }}\\	\hline
&\multicolumn{4}{|l|}{\scriptsize Warm optics}&&\multicolumn{4}{l|}{\scriptsize $\pm 17$ \%} \\ 
\cline{2-5}\cline{7-10}
&&\multicolumn{3}{|l|}{\scriptsize Fiber transmission}&&&\multicolumn{3}{l|}{\scriptsize $\pm 51$ \%} \\
\cline{3-5}\cline{8-10}
&&&\multicolumn{2}{|l|}{\scriptsize Fiber coupling}&&&&\multicolumn{2}{l|}{\scriptsize $\pm 81$ \%} \\
\cline{4-5}\cline{9-10}
&&&\multicolumn{2}{|l|}{\scriptsize Throughput of the fiber}&&&&\multicolumn{2}{l|}{\scriptsize $\pm 66$ \%} \\
\cline{4-5}\cline{9-10}
&&&\multicolumn{2}{|l|}{\scriptsize Coupling degradation factor}&&&&\multicolumn{2}{l|}{\scriptsize $\pm 95$ \%} \\
\cline{4-5}\cline{9-10}
&&&&\scriptsize Fiber head shift&&&&&\scriptsize $\pm0.3 \mu$m \\
\cline{4-5}\cline{9-10}
&&&&\scriptsize Defocus&&&&&\scriptsize $\pm4 \mu$m \\
\cline{4-5}\cline{9-10}
&&&&\scriptsize Fiber angular inclination&&&&&\scriptsize $\pm0.5^\circ$ \\
\cline{4-5}\cline{9-10}
&&&&\scriptsize Optical quality&&&&&\scriptsize $\lambda$/10 rms $@$ 633 nm \\
\cline{4-5}\cline{9-10}
&&&&\scriptsize Polarization control&&&&&\scriptsize $\pm 3^\circ $ \\
\cline{3-5}\cline{8-10}
&&\multicolumn{3}{|l|}{\scriptsize Warm optics optical element transmission}&&&\multicolumn{3}{l|}{\scriptsize $\pm 33$ \%} \\
\cline{2-5}\cline{7-10}
&\multicolumn{4}{|l|}{\scriptsize SPG}&&\multicolumn{4}{l|}{\scriptsize $\pm 39$ \%} \\ 
\cline{2-5}\cline{7-10}
&\multicolumn{4}{|l|}{\scriptsize DET}&&\multicolumn{4}{l|}{\scriptsize $\pm 62$ \%} \\ \hline\hline
\multicolumn{5}{|l|}{\scriptsize \textbf{Static chromatic OPD}}& \multicolumn{5}{l|}{\scriptsize \textbf{$\leq 0.1\mu$m inside the spectral bands}}\\	\hline
&\multicolumn{4}{|l|}{\scriptsize Relative glass thickness and fiber length}&&\multicolumn{4}{l|}{\scriptsize $\approx\pm 500 \mu$m shared out all the transmissive elements} \\ 
\cline{2-5}\cline{7-10}
&\multicolumn{4}{|l|}{\scriptsize Optical quality of each element}&&\multicolumn{4}{l|}{\scriptsize $\lambda$/5 PTV $@$ 633 nm } \\ \hline\hline
\multicolumn{5}{|l|}{\scriptsize \textbf{Dynamic chromatic OPD}}& \multicolumn{5}{l|}{\scriptsize \textbf{$\leq$ photon noise during 1 minute}}\\	\hline 
&\multicolumn{4}{|l|}{\scriptsize Beam angular deviation at the exit of prismatic optics}&&\multicolumn{4}{l|}{\scriptsize $\pm1''$ to $\pm3''$} \\
\cline{2-5}\cline{7-10} 
&\multicolumn{4}{|l|}{\scriptsize Relative glass thickness and fiber length}&&\multicolumn{4}{l|}{\scriptsize $\approx\pm 500 \mu$m shared out all the transmissive elements} \\ 
\hline \hline
\multicolumn{5}{|l|}{\scriptsize \textbf{Contrast calibration (P2VM)}}& \multicolumn{5}{l|}{\scriptsize \textbf{3$\sigma_{V}$=0.01}}\\	\hline
&\multicolumn{4}{|l|}{\scriptsize Generated contrast of point-like source}&&\multicolumn{4}{l|}{\scriptsize $>$75\%} \\ 
\cline{2-5}\cline{7-10}
&\multicolumn{4}{|l|}{\scriptsize Contrast knowledge accuracy}&&\multicolumn{4}{l|}{\scriptsize $10^{-3}$} \\  \hline\hline
\multicolumn{5}{|l|}{\scriptsize \textbf{Dimensioning}}& \multicolumn{5}{l|}{}\\	\hline
&\multicolumn{4}{|l|}{\scriptsize VLTI AMBER interface}&&\multicolumn{4}{l|}{} \\
\cline{2-5}\cline{7-10}
&&\multicolumn{3}{|l|}{\scriptsize Beam size}&&&\multicolumn{3}{l|}{\scriptsize 18 mm} \\
\cline{3-5}\cline{8-10}
&&\multicolumn{3}{|l|}{\scriptsize Beam height adjustment accuracy}&&&\multicolumn{3}{l|}{\scriptsize $\pm$0.3 mm} \\
\cline{3-5}\cline{8-10}
&&\multicolumn{3}{|l|}{\scriptsize Beam separation adjustment accuracy}&&&\multicolumn{3}{l|}{\scriptsize $\pm$0.3 mm} \\ \hline
&\multicolumn{4}{|l|}{\scriptsize SPG warm optics: interface}&&\multicolumn{4}{l|}{} \\
\cline{2-5}\cline{7-10}
&&\multicolumn{3}{|l|}{\scriptsize Beam size and separations (no anamorphosis)}&&&\multicolumn{3}{l|}{\scriptsize See Fig.~\ref{pupil}} \\
\cline{3-5}\cline{8-10}
&&\multicolumn{3}{|l|}{\scriptsize Anamorphosis factor}&&&\multicolumn{3}{l|}{\scriptsize 15.6} \\
\cline{3-5}\cline{8-10}
&&\multicolumn{3}{|l|}{\scriptsize Pupil transverse location accuracy}&&&\multicolumn{3}{l|}{\scriptsize $\pm250 \mu$m in the anamorphosis direction} \\
&&\multicolumn{3}{|l|}{}&&&\multicolumn{3}{l|}{\scriptsize $\pm16 \mu$m in the other one} \\
\cline{3-5}\cline{8-10}
&&\multicolumn{3}{|l|}{\scriptsize Image transverse location accuracy}&&&\multicolumn{3}{l|}{\scriptsize $\pm5 \mu$m in the anamorphosis direction} \\\hline
		\end{tabular}
\end{center}
\label{detail}
\end{table*}

\subsubsection{Error budget on the fiber coupling}
This paragraph gives a list of the errors that can contribute to the degradation of the fiber coupling $C_{eff}$. To derive this error budget, analytical simulations were performed \citep{escarrat}. These results were confirmed with the ray tracing tool Zemax in which the optical injection of light inside a fiber can be estimated \citep{wagner}. \newline
Table \ref{spec_fib} shows the error budget on the fiber coupling in terms of fiber axis inclination respective to the incident optical axis (tilt), lateral shift of the fiber head, and defocus compared with the localization of the injection optical focal point. Coupling efficiencies without misalignments were also estimated after modification of the ratio F/D, where F is the injection optical focal length and D the pupil diameter. 
\begin{table}[h]
	\caption{Error budget on the coupling efficiency. F = focal length of the injection optics, D = pupil diameter.}
		\begin{tabular}{c c c c c} \hline\hline
	&\scriptsize Deviation &\scriptsize 1.10 $\mu$m&\scriptsize	1.65 $\mu$m&\scriptsize	2.20 $\mu$m\\	
	&\scriptsize from optimum&\scriptsize &\scriptsize	&\scriptsize	\\	\hline
\scriptsize Tilt&\scriptsize	0.5$^\circ$ &\scriptsize	$>$ 0.99&\scriptsize	$>$ 0.99&\scriptsize	$>$ 0.99\\
\scriptsize Fiber head shift&\scriptsize	0.3 $\mu$m&\scriptsize	$>$ 0.98	&\scriptsize$>$ 0.99	&\scriptsize$>$ 0.99\\
\scriptsize Defocus	&\scriptsize4 $\mu$m	&\scriptsize$>$ 0.99&\scriptsize	$>$ 0.99&\scriptsize	$>$ 0.99\\
\scriptsize F/D&\scriptsize	0.93&\scriptsize	$>$ 0.99&\scriptsize	$>$ 0.99&\scriptsize	$>$ 0.99\\\hline
\scriptsize \textbf{Total}	&\scriptsize&\scriptsize	$>$ 0.95&\scriptsize	$>$ 0.96	&\scriptsize$>$ 0.96\\\hline
		\end{tabular}
	\label{spec_fib}
\end{table}
This ratio depends on the numerical aperture. The error budget on fiber coupling requirements was then compared to VLTI performance to validate the hypotheses.
It was the input for the definition of the degrees of freedom necessary on each optical element located at the entrance of the fibers and to achieve their detailed tolerance analysis. 
\subsubsection{Contrast error budget}
Results on contrast-error budget in the K-band (priority of AMBER) with a 35 spectral resolution (most demanding case) are shown in Table \ref{contrast}. 
This error budget concerns differential errors between interferometric beams. A degradation factor $\rho$ is such that the measured instrumental contrast $V_{inst}$ is equal to the ideal contrast value $V_{i}$ times this factor ($V_{inst}= \rho \; V_{i}$). It was assumed in a first approximation that the errors were independent. 
Some of the requirements concerning static errors were analytically derived. Others requiring dynamical analysis needed some simulation. The latter considered a Gaussian wavefront going through a simple 2-telescope interferometer using the AMBER spectral bandwidths and resolutions. Four pixels were considered to sample the fringes. The considered pupil dimensions after anamorphosis in the horizontal direction were 40x2.6 mm in K, 30x1.9 mm in H, and 20x1.5 mm in J. The simulation estimated the normalized optical transfer function (OTF), which consists of three peaks from which the fringe contrast can be evaluated: two high frequency peaks containing the coherent energy, and one low frequency peak containing the incoherent energy (Fig.~\ref{otf}). The fringe contrast was then defined as the ratio of the coherent to the incoherent energies.
More details on the error definition are given in the Appendix.
The results listed in Table \ref{contrast} were used to define the degrees of freedom necessary on each optical element located after the exit of the fibers and to achieve their detailed tolerance analysis. 
\begin{table} [h]
	\caption{Contrast degradation factors and allocations derived in the K-band with a 35 spectral resolution. $l_{c}$ is the coherence length and $Airy$ the Airy disk size.}
\begin{center}
		\begin{tabular}{c c c c} \hline\hline
 \scriptsize \textbf{Allocations}	&\scriptsize  \textbf{$\rho$}  &\scriptsize 	\textbf{Contrast} &\scriptsize  \textbf{Requirement}\\
&\scriptsize &\scriptsize  \textbf{noise} &\scriptsize  \textbf{rms}	\\	\hline
\scriptsize OPD equality &\scriptsize 	0.995	 &\scriptsize 	&\scriptsize  $l_{c}$/19 (4 $\mu$m)\\
\scriptsize OPD vibration (rms)	&\scriptsize &\scriptsize 	10$^{-3}$	&\scriptsize   $\lambda$/140 (16 nm)\\
\scriptsize Image overlapping	&\scriptsize  0.99 &\scriptsize 	&\scriptsize 	$Airy$/10 (3")\\
\scriptsize Overlapping jitters &\scriptsize  &\scriptsize  10$^{-3}$	&\scriptsize  $Airy$/12 (3")\\
\scriptsize Wavefront quality &\scriptsize 	0.95	 &\scriptsize  &\scriptsize 	  $\lambda$/16 (140 nm)\\
\scriptsize Defocus	&\scriptsize  0.99	&\scriptsize &\scriptsize  	$\approx$ 10 mm\\
\scriptsize Phase delay	&\scriptsize  0.99&\scriptsize &\scriptsize 		17$^\circ$\\
\scriptsize Differential rotation of&\scriptsize &\scriptsize &\scriptsize \\
\scriptsize polarization frames&\scriptsize  	0.99	&\scriptsize &\scriptsize 	16$^\circ$\\
\scriptsize Unequal intensity	&\scriptsize  0.999	&\scriptsize &\scriptsize 	I$_{1}$/I$_{2}$ $\approx$ 0.9\\\hline
\scriptsize \textbf{Total}	&\scriptsize   $\approx$ 0.89&\scriptsize  	$\approx$ $10^{-3}$ &\scriptsize 	\\\hline
		\end{tabular}
		\end{center}
	\label{contrast}
	\end{table}
	
\subsection{Optical study}
\begin{figure} [ht]
\includegraphics[width=85mm]{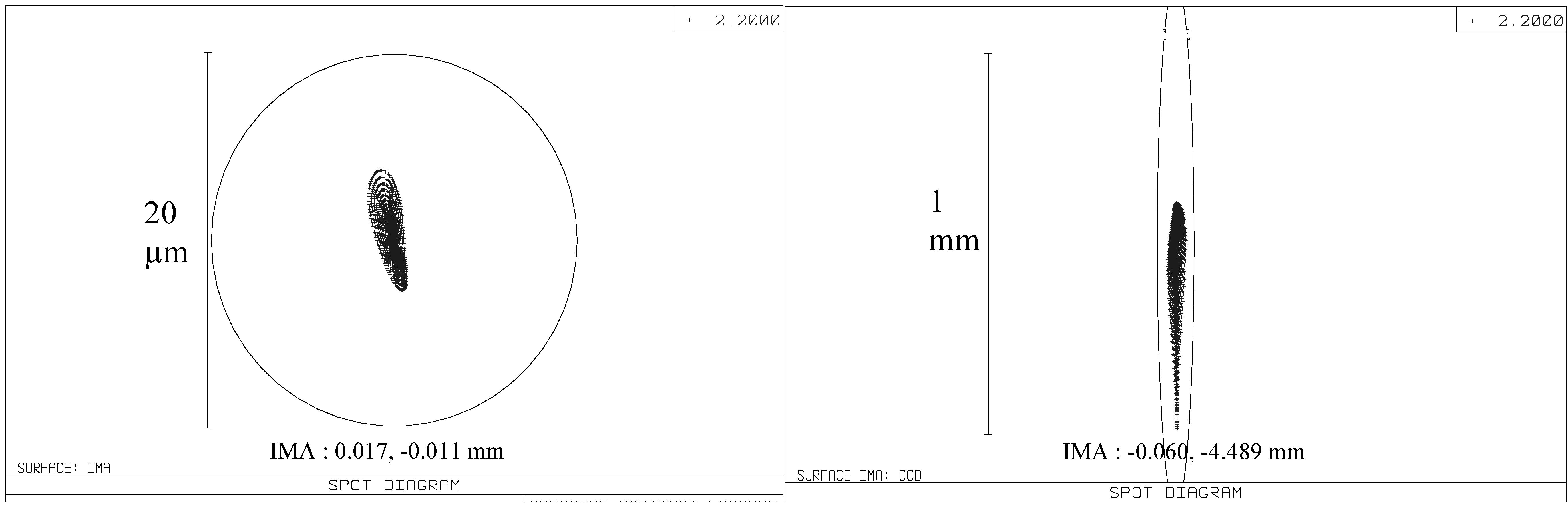}
\caption{Left: spot diagram at the SFK fiber entrance using the light from the VLTI. The coupling loss ratio is about 2.6\%. Right: spot diagram at the spectrograph entrance in the K-band. The associated Strehl ratio is 0.92. This optical quality leads to a differential WF rms error of $\lambda$/15.5 @ 2.2 $\mu$m, which corresponds to a 0.95 instrumental contrast degradation factor.}
\label{spot}
\end{figure}

The optical configuration performance was assessed by the study of spot diagrams at the fiber entrance and by the estimation of the optical quality at the entrance of the spectrograph (Fig.~\ref{spot}). These respectively reflect the expected maximum fiber coupling efficiency (Table \ref{spec_fib}) and the expected maximum contrast (Table \ref{contrast}). 
The ray tracing optical tool Zemax was used to test the tolerance of all elements and surfaces. It is able to estimate a fiber coupling efficiency and to perform the tolerance analysis of the elements located before the fiber entrance such that the coupling stays very near a maximum value previously defined. Zemax is not able to estimate a fringe contrast for a configuration such as that of AMBER, but provides the wavefront (WF) optical quality taken into account in the contrast budget.

\subsubsection{Estimation of the optical throughput from the optical study}
Concerning the AMBER warm optics:
the expected coupling efficiency degradation factor estimated from the tolerance analysis is 96\% in K, including the VLTI WF errors. This, in addition to the throughput measurements provided by the optics manufacturers and to the coupling efficiency, leads to a global optical throughput $t_{A} C_{eff}$ of 8\% to 12\%, depending on the wavelength (VLTI included).
Including the spectrograph and the detector efficiency in the estimations lead to a global throughput of 2\% to 4\% meeting the specifications.

\subsubsection{Estimation of the contrast from the optical study}
At the entrance of the spectrograph, the expected optical quality of one beam is expressed in terms of differential WF rms error: $\lambda$/15.5 @ 2.2 $\mu$m, which corresponds to a contrast degradation factor of 0.95 in K. This represents the instrumental factor to be applied to the fringe visibility, due to the WF quality of the AMBER warm optics located at the exit of the fibers.
The contrast budget of the warm optics taking into account additional parameters such as achromatic OPD errors, differential chromatic OPD, global tilt errors, defocus, differential direction of polarization, and unequal intensity, is about 0.90 in K.

\begin{figure} [ht]
\begin{center}
\includegraphics[width=60mm]{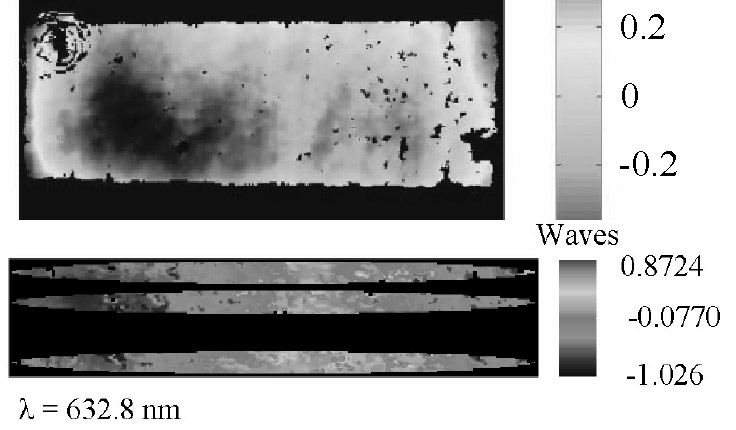}
\includegraphics[width=50mm]{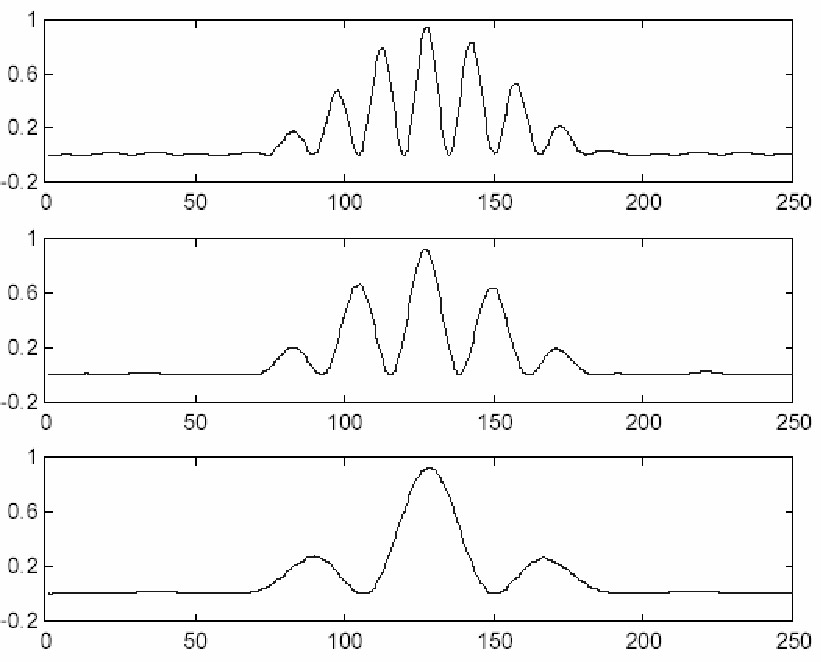}
\end{center}
\caption{Above: Residual WF aberration at liquid nitrogen temperature as measured with external optical elements (45x16 mm surface). Center: Three pupil masks superimposed to the measured wavefront. Below: Simulated PSF at 2200 nm for the combinations of two apertures at different separation.}
\label{spg_wf}
\end{figure}

The wavefront aberration of the optical system from the input slit
to the detector plane has a measured PTV value of 2.8 fringes $@$ 633 nm
and a RMS of about 200 nm (on the largest pupil area the system exploits).
This value is well inside the error budget for SPG. The contribution to
the fringe contrast degradation factor of AMBER coming from SPG is estimated to be greater
than 98$\%$. Figure~\ref{spg_wf} shows the residual wavefront aberration at liquid nitrogen temperature as measured with external optical elements, independently of the warm optics. The surface size is 45x16 mm. The pupil mask was then superimposed to the measured WF to simulate the final configuration and to estimate the associated PSF and fringe contrast. The evaluation of the AMBER instrumental contrast of 88\% from the optical study is higher than the specifications.

\subsection{Stability}
To get very accurate phase and contrast measurements, the interferometer has to be as stable as possible. 
To observe fine structures in the disks and jets around a sufficient number of young stellar objects, it is necessary to obtain a requirement of 0.01 \citep{malbet} on the instrument contrast stability during 5 minutes, the initial estimation of the calibration cycle.
There are several sources of instabilities  that need to be taken into account including: temperature gradients, vibrations, and micro-turbulence.

\subsubsection{Temperature gradients}
Temperature gradients can occur when the temperature is not correctly monitored or when heating (active motors, people, sources on, ...) or cooling (cooled dewar) sources are present close to the instrument.
The global dissipation in the laboratory was measured \citep{puech}: +50 W for motors and electronics and -20 W for the cryostat. These internal fluxes are not a problem so far, but the effects on the highest goals still have to be analyzed.

The measured temperature gradient through the AMBER table, from the CAU to the spatial filter J, is currently 0.03$^\circ$K. It is stable over one night if there are no interventions in or near AMBER. The fastest fluctuations are lower than the current 0.01$^\circ$K monitoring accuracy. This corresponds to predicted fluctuations of the differential visibility and phase smaller than 0.001~rad \citep{martin}. This is confirmed by our current measurements, which are currently limited to a few 0.001~rad accuracy by the atmosphere. If the external conditions allow us to achieve the highest accuracy of 10$^{-4}$ rad necessary for exoplanets spectroscopy (while using the ESO fringe sensor FINITO for example), the current thermal stability of AMBER of 10$^{-3}$ rad will prove to be insufficient. \citet{martin} showed that a possible solution to reach such a goal would be to thermally insulate the AMBER fibers.

\subsubsection{Vibrations}
Vibrations, inducing OPD fluctuations, can occur inside the instrument when internal sources of vibration exist (motors, ventilation, nitrogen bubbling, water cooling devices,...) and/or when external vibrations are transmitted to the instrument via the floor.\newline
Optical tables can also act as cavity resonators according to their damping capabilities. \newline
To address this, specific considerations (pneumatic feet, water chiller, isolation of water cooling pipes, use of screw-type pumps, placing all vibration generating equipment on soft pads) and an estimation of eigen frequencies were performed.

\subsection{Measurements in laboratory}
The performance of AMBER measured in laboratory \citep{rouss} for the Preliminary Acceptance in Europe (PAE) showed that the instrument reaches the requirement and approaches the goals.
Although absolute photometric measurements could not be performed in laboratory as no dedicated bench was available, relative measurements gave a throughput of 5\% in K, 13\% in H, and 10\% in J. As regards to these values, limiting magnitudes of 11 could be reached.
The instrumental contrasts measured in K with the CAU were: 0.78 at Low Resolution (LR), 0.87 at Medium Resolution (MR), and 0.83 at High Resolution (HR). Let us not forget that the CAU sources are slightly resolved by AMBER. These values were taken in the presence of the polarizers. Removing them has the consequence to reduce the contrasts by up to few tens of \%. 
An accuracy of 1\% on the CAU contrast at 2.2$~\mu$m was measured, meeting the specification.
 The specifications of 10$^{-2}$ over 5 minutes on contrast stability and of a few 10$^{-3}$~rad over 1 minute on differential phase stability are met.
Additional measurements were performed with the CAU during the alignment integration and verification (AIV) phase at Paranal \citep{robb}. The instrumental contrasts in H and J were very similar to those in K.

\subsection{Results of commissionings}
This paragraph summarizes the situation of AMBER after its integration (February-March 2004) and its first two commissioning runs (May and October 2004). More details are given in \citep{petrovd}. The 3rd commissioning took place in February 2006, its analysis is in progress and will be reported in http://www-laog.obs.ujf-grenoble.fr/amber/.

AMBER is fully operational in 2 telescopes and 3 telescopes modes.
It has been commissioned in MR in the K-band and in LR in the K- and H-bands. For AMBER alone, the contrast transfer function is in the 65\%-85\% range depending on modes. The contrast, differential phase, and phase closure stability are of a few 10$^{-3}$.

The flux collecting efficiency of the AMBER/VLTI system was higher than expected, in spite of the absence of the VLTI image sensor in the focal laboratory. This corresponds to a limiting magnitude higher than 11 in LR and higher than 7 in MR (without fringe tracker).
So far we have been able to acquire objects and to detect fringes up to K=9 in LR and K=7 in MR. In MR, this is very close to the theoretical limit. In LR, the object acquisition was inefficient due to the absence of the ESO infrared image sensor in the focal laboratory at the time of the first commissioning runs. The implementation of the image sensor has dramatically improved the acquisition speed and the longer term stability of the injection, but has not radically changed the fiber injection efficiency in the minutes following the source acquisition.

The data processing of AMBER is operational. The delivered routines can be used as black boxes for applications in the 1\% accuracy range, given that the appropriate detector calibration procedure has been used.
In MR, there is no operational difference observing with 2 or 3 telescopes. In LR, with the appropriate VLTI OPD model (controlling the delay line stroke in each interferometric arm) and with a fringe search simultaneous for all baselines, the operation with 3T is exactly as easy as the operation with 2T.

Almost all the data recorded with AMBER on the UTs is affected by a piston vibration problem coming mainly from the UT coude train and generated by several external sources such as the fans of the electronics cabinets of the MACAO units or the cryo-coolers of the other VLT instruments. So far, we have been able to reduce data only for K=5 in MR and K=8.5 in LR. Within these magnitudes, the accuracy of the AMBER/VLTI transfer function is better than 1\%, even with the vibrations, achieved in a few minutes of observations. The improvement of the MACAO system is now achieved.

The HR mode in K has been commissioned down to K=2 and is currently offered.
The implementation of the FINITO fringe tracker with expected limiting magnitudes of about 7 with the ATs and 10 with the UTs (once the vibration problem is solved) will allow operating in all the spectral resolution modes at this limiting magnitudes.
No commissioning of the J-band has been made. This commissioning needed the service of the ADC, which has just been successfully tested on sky.

The first results on the differential phase are described by \citet{vannierb}: they show the RMS instrumental stability of the differential phase over 60s in the three spectral bands and in the low resolution mode in the order of 10$^{-3}$~rad on two baselines and about 10$^{-4}$~rad on the third one. Some solutions (technical or concerning the data treatment) are under investigation to improve or correct the stability on the two lowest baselines.
\newline

The last commissioning with the ATs (July 2006) measured instrumental contrast in the 70\%-90\% range (including the atmosphere, the VLTI, and AMBER, this contrast being corrected from the source estimated visibility). Similar values should be reached with the UTs when the vibrations are eliminated or actively controlled. The atmosphere alone is supposed to produce about 10\% in contrast loss and the VLTI was initially specified to produce an additional loss lower than 10\%.
With the currently measured transmission and the atmosphere+VLTI+AMBER fringe contrast, the limiting magnitude can be extrapolated to be between K=9 and K=10 \citep{petrovd} in the 20\% best conditions. If the vibrations with the UTs are reduced to the level of the ATs and initially specified for the VLTI, these limiting magnitudes will be of the order of K=11 with the UTs \citep{petrovd}. AMBER is therefore fully compliant with its initial sensibility specifications.

\section{Conclusion}
To achieve the ambitious astrophysical program of AMBER, it is necessary to obtain visibility measurements with an accuracy better than 10$^{-2}$ (goal: 10$^{-3}$) on sources of magnitude K relatively smaller than 11 and to reach an instrumental stability of the differential phase of 10$^{-3}$ rad (goal: 10$^{-4}$ rad) over one minute.
It was shown that these general specifications are driven by the necessity for the instrument to provide interferometric requirements that consist mainly of contrast level and accuracy, optical throughput and stability. To assess the AMBER performance, a tolerance procedure was defined: with the scientific specifications, the interferometric specifications were determined, in terms of error budgets for the fiber coupling degradation factor due to misalignments and for the instrumental contrast. Both allowed us to define the degrees of freedom of each optical element, respectively located before and after the optical fibers, and represented inputs for the detailed tolerance analysis performed with a ray-tracing tool.

The complete tolerance study, confirmed by the observations, assessed the feasibility of the requirements, i.e., an overall instrumental contrast level higher than 80\% in the K-band and the 11th K-magnitude with the lowest resolution.
After separate integration and tests in the institutes in charge of the warm optics \citep{roba}, SPG \citep{lisi}, and DET, the whole instrument was integrated and tested in LAOG in 2003. 
AMBER successfully passed the Preliminary Acceptance in Europe in November 2003, resulting in the validation of the instrument laboratory performance, of the compliance with the initial scientific specifications \citep{rouss}, and of the acceptance of ESO for AMBER to be part of the VLTI. After the transportation of the instrument to Paranal, Chile in January 2004, the Assembly Integration and Verification phase occurred mid-March with a successful first fringe observation of bright stars with the VLTI siderostats \citep{robb}. 

Since then four commissioning runs have been performed, strengthening the performance. The astrophysical results presented in this issue of A\&A is the best evidence that AMBER is today a very powerful tool allowing unequalled astrophysical knowledge. More results are expected once VLTI has completely solved vibration problems and installed the fringe sensor.
\begin{acknowledgements}
 The AMBER project\footnote{The structure and members of the AMBER Consortium can be found in the website: \texttt{http://amber.obs.ujf-grenoble.fr}} has been founded by the French Centre National de la Recherche Scientifique (CNRS), the Max Planck Institute f\"{u}r Radioastronomie (MPIfR) in Bonn, the Osservatorio Astrofisico di Arcetri (OAA) in Firenze, the French Region "Provence Alpes C\^{o}te D'Azur" and  the European Southern Observatory (ESO). The CNRS funding has been made through the Institut National des Sciences de l'Univers (INSU) and its Programmes Nationaux (ASHRA, PNPS, PNP).

The OAA co-authors acknowledge partial support from MIUR grants to the Arcetri Observatory: \emph{A LBT interferometric arm, and analysis of VLTI interferometric data} and \emph{From Stars to Planets: accretion, disk evolution and planet formation} and from INAF grants to the Arcetri Observatory \emph{Stellar and Extragalactic Astrophysics with Optical Interferometry}. C. Gil work was supported in part by the Funda\c{c}\~ao para a Ci\^encia e a Tecnologia through project POCTI/CTE-AST/55691/2004 from POCTI, with funds from the European program FEDER.

The preparation and interpretation of AMBER observations benefit from the tools developed by the Jean-Marie Mariotti Center for optical interferometry JMMC\footnote{The JMMC is a center providing software tools for optical interferometry described at the website: \texttt{http://www.mariotti.fr/}} and from the databases of the Centre de Données Stellaires (CDS) and of the and the Smithsonian/NASA Astrophysics Data System (ADS).

We would like to thank the successive directors of INSU/CNRS, G. Debouzy, F. Casoli and A.-M. Lagrange, the ASHRA president P. Lena and the ESO VLT-Program-Manager M. Tarenghi for their crucial help in setting up and supporting the Consortium. We also thank S. Bensammar, V. Coud\'{e} du Foresto, G. Perrin for their advice in defining the AMBER concept and in setting up the Consortium.

We are very grateful to the ESO staff in Garching and Paranal for their help in the design and the commissioning of AMBER.

Key mechanical design and manufacturing has been provided by the Division Technique de l'INSU and by the mechanical workshops of the Observatoire de Bordeaux, Observatoire de la C\^{o}te d'Azur and Universit\'{e} de Nice.
\end{acknowledgements}

\begin{appendix}
\section{Details on the contrast error budget}
\subsection{OPD equality} 
The fringe contrast degradation factor $\rho_{opd}$ due to an OPD $\delta$ between the two arms of the interferometer is:
\begin{equation}
\rho_{opd} = \frac{sin(\pi \delta / l_{c})}{\pi \delta / l_{c}},
\end{equation}
where $l_{c}$ is the coherence length equal to 
$\lambda^{2}$ / $\Delta\lambda$. This is considering a constant spectrum for the source on the observation bandwidth $\Delta\lambda$.
\subsection{OPD vibration (rms)}
Some piston or rms OPD variation effects introduce a complex term in the pupil transmission terms that can be written: $P_{1,2}(x,y)=P_{0}(x,y).e^{i.\phi_{1,2}}$.
The associated contrast degradation factor is \citep{rob} $\rho_{vib} = < e^{i.\Delta\phi} >$, where $\Delta\phi$ is the instantaneous differential phase and $<>$ represents the average during the integration time. From \citep{roddier}: 
\begin{equation}
\rho_{vib} = e^{-\sigma^2/2} \approx 1 - \sigma^2/2 ,
\end{equation}
with $\sigma^2$ the phase variance, the approximation being valid for very small variations.\newline
As an example, considering a rms OPD variation of $\lambda$/140, the phase variance is then $\sigma^2=(2\pi/140)^2$ leading to a degradation factor of 0.999 and a contrast noise $1-\rho_{vib}$ of $10^{-3}$.
Simulations gave the same results and also showed that the degradation is independent of the spectral resolution.
\subsection{Static image overlapping errors}
Static tracking errors appear when a differential tilt between the two combined wavefronts exists,
inducing a non-perfect overlapping of the two Airy disks.
The phase $\phi$ in the wavefront, as a function of a tilt ($\alpha$, $\beta$), is given by:
$\phi(x,y)=e^{i \frac{2\pi}{\lambda} (\alpha x + \beta y)}$,
where $x, y$ are the spatial coordinates and $\alpha$, $\beta$ are the inclinations
of the wavefront.
The wavefront becomes:\newline
$\Psi(x,y) = P_{1}(x,y)~e^{i\frac{2\pi}{\lambda} (\alpha _{1} x + \beta _{1} y)}+P_{2}(x,y)~e^{i\frac{2\pi}{\lambda} (\alpha _{2} x + \beta _{2} y)} $
where $P_{1,2}(x,y)=P_{0}(x,y)\otimes \delta(x \pm \frac{B}{2})$, with $P_{0}$ the pupil transmission,
$\otimes$ the convolution operator, $\delta$ the Dirac's function, and $B$ the beam separation.
The associated contrast degradation factor  $\rho_{tilt}$  is \citep{rob}:
\begin{equation}
\rho_{tilt} = ~\frac {\widetilde{P}_{0}(u-\frac{\alpha_{1}}{\lambda},v-\frac{\beta_{1}}{\lambda})~
\widetilde{P}_{0}^{*}(u-\frac{\alpha_{2}}{\lambda},v-\frac{\beta_{2}}{\lambda})}
{\left| \widetilde{P}_{0}(u-\frac{\alpha_{1}}{\lambda},v-\frac{\beta_{1}}{\lambda})\right|^{2} +
\left| \widetilde{P}_{0}^{*}(u-\frac{\alpha_{2}}{\lambda},v-\frac{\beta_{2}}{\lambda})\right|^{2}},%
\end{equation}
where $\sim$ denotes the Fourier Transform and * the complex conjugate.
This term can be analytically expressed considering a co-axial acquisition of the fringes. $\rho_{tilt}$ can then be approximated by $\rho_{tilt} = ~\frac {2J_{1}(Z)}{Z}$,
where $J_{1}$ is the first order Bessel function of the first kind and 
$Z = \frac{2 \pi R w}{\lambda}$, with 
$w = \sqrt{(\alpha_{1}-\alpha_{2})^{2}+(\beta_{1}-\beta_{2})^{2}}$ and $R$ 
the circular pupil radius.
For a better estimation, simulations using the AMBER multi-axial configuration were performed to assess the requirements. 
The requirements are equivalent in any spectral resolution.
\subsection{Overlapping jitters (rms)}
Assuming a Gaussian distribution for the pointing errors, the mean contrast was evaluated as
a function of the overlapping error standard deviations.
\subsection{Wavefront quality (rms)}
Assuming a Gaussian distribution for the phase defects originated in departures from an ideal shape of the optical surface, the mean contrast is evaluated as a function of the error standard deviations.
Simulations show that the results are strongly similar in every spectral bandwidth.
The results are quite similar to (a little more optimistic than) the approximation usually used for the degradation factor  $\rho_{wfe}= e^{-\sigma^2/2}$ where $\sigma$ is the rms differential wavefront error (WFE) in radians.
\subsection{Defocus}
The error in focusing was introduced in the simulations via the associated
Zernike polynomial. Results show that the contrast loss depends on the spectral bandwidth.
\subsection{Polarization: phase delay and differential rotation}
Polarization effects were described by \citet{roussa} and we just give the analytical results here.
A contrast degradation can be induced by a differential instrumental polarization. Each direction of polarization, s and p, produces an independent interference pattern. Both add incoherently. A differential phase delay of $\phi_s-\phi_p$ between the interference patterns introduces a fringe shift $\phi'= \frac{\phi_{s}+\phi_{p}}{2}$ and a visibility degradation factor $\rho_{pol}$:\begin{equation}
\rho_{pol} = cos(\frac{\phi_{s}-\phi_{p}}{2}).
\end{equation}
Now, if there exists an angle $\theta_{12}$ between the reference frame between the two arms
of the interferometer, the contrast degradation factor $\rho_{rot}$ is:
\begin{equation}
\rho_{rot} = \frac{2 \left|cos(\theta_{12})\right|}{1+cos^2(\theta_{12})}.
\label{eq12}
\end{equation}
\end{appendix}

\end{document}